\documentclass[fleqn,usenatbib]{mnras}

\usepackage{newtxtext,newtxmath}

\usepackage[T1]{fontenc}
\usepackage{ae,aecompl}

\usepackage{graphicx}	
\usepackage{amsmath}	
\usepackage{amssymb}	
\usepackage{float}
\usepackage{tabularx}

\newcommand\eagle{{\sc EAGLE}}       
\newcommand\gadget{{\sc GADGET}}       
\newcommand\subfind{{\sc subfind}}       
\newcommand\velociraptor{{\sc VELOCIraptor}}       
\newcommand\treefrog{{\sc TreeFrog}} 
\newcommand\chumm{{\sc chumm}} 

\title[Halo-scale mass assembly in \eagle]{The impact of stellar and AGN feedback on halo-scale baryonic and dark matter accretion in the \eagle\ simulations}

\author[R. J. Wright et al.]{Ruby J. Wright\thanks{E-mail: ruby.wright@icrar.org}$^{ 1,2}$, Claudia del P. Lagos$^{1,2}$, Chris Power$^{1,2}$, Peter D. Mitchell$^{3}$ 
\\
$^{1}$International Centre for Radio Astronomy Research (ICRAR), University of Western Australia, Crawley, WA 6009, Australia\\
$^{2}$ARC Centre of Excellence for All Sky Astrophysics in 3 Dimensions (ASTRO 3D)\\
$^{3}$Leiden Observatory, Leiden University, P.O. Box 9513, 2300 RA Leiden, the Netherlands\\}

\date{Accepted XXX. Received YYY; in original form ZZZ}

\pubyear{2020}

\begin{document}
\label{firstpage}
\pagerange{\pageref{firstpage}--\pageref{lastpage}}
\maketitle

\begin{abstract}
We use the \eagle\ suite of hydrodynamical simulations to analyse accretion rates (and the breakdown of their constituent channels) onto haloes over cosmic time, comparing the behaviour of baryons and dark matter (DM). We also investigate the influence of sub-grid baryon physics on halo-scale inflow, specifically the consequences of modelling radiative cooling, as well as feedback from stars and active galactic nuclei (AGN). We find that variations in  halo baryon fractions at fixed mass (particularly their circum-galactic medium gas content) are very well correlated with variations in the baryon fraction of accreting matter, which we show to be heavily suppressed by stellar feedback in low-mass haloes, $M_{\rm halo}\lesssim10^{11.5}M_{\odot}$. Breaking down accretion rates into first infall, recycled, transfer and merger components, we show that baryons are much more likely to be smoothly accreted than to have originated from mergers when compared to DM, finding (averaged across halo mass) a merger contribution of $\approx6\%$ for baryons, and $\approx15\%$ for DM at $z\approx0$. We also show that the breakdown of inflow into different channels is strongly dependent on sub-grid physics, particularly the contribution of recycled accretion (accreting matter that has been previously ejected from progenitor haloes). Our findings highlight the dual role that baryonic feedback plays in regulating the evolution of galaxies and haloes: by (i) directly removing gas from haloes, and (ii) suppressing gas inflow to haloes. 

\end{abstract}




\section{Introduction}\label{sec:introduction}
\subsection{Motivations \& the current paradigm of mass assembly}
Cosmological simulations focused on the physics of galaxy formation and evolution have proven to be exceptionally powerful predictive tools in extragalactic astrophysics. Hydrodynamical simulations and semi-analytic models are developing in parallel, acting to support observational surveys in constraining the complex baryonic physics that takes place within galaxies, and their consequent observable properties \citep{Somerville2015}.

 Semi-analytic models are one of the most frequently used tools to investigate the physics behind galaxy formation and evolution on large scales. This is largely due to their ability to explore vast regions of parameter space, and produce sizeable galaxy populations without prohibitive computational cost. Many SAMs (e.g. {\sc GALFORM}: \citealt{Cole2000,Lacey2016}, {\sc l-galaxies}: \citealt{Henriques2015}, and {\sc Shark}: \citealt{Lagos2018b}) build galaxy populations based on dark matter (DM) only simulations, and must accordingly infer the growth of baryons in galaxies based on the DM growth of the host halo. The behaviour of baryons is normally assumed to trace exactly that of DM, with a mass fraction of $f_{\rm b}=\frac{\Omega_{\rm b}}{\Omega_{\rm m}}\approx 0.165$, defined by the parameters of a concordance $\Lambda$ Cold DM ($\Lambda$CDM) cosmology \citep{Planck2018}. 

In a no-feedback regime, \citet{Crain2007} show this assumption to be valid in cosmological hydrodynamical simulations. They found that the baryon fraction of haloes within the virial radius is $(\approx90\pm6)\%\times f_{\rm b}$, independent of halo mass and redshift. This said, the assumption that the baryonic mass traces DM has not been extensively explored in modern simulations. The aim of our work is to approach this multi-faceted problem by making use of state-of-the-art cosmological hydrodynamical simulations ({\it including feedback physics}) to investigate gas and DM inflow, with the eventual goal of applying our results to better inform SAMs. 

In the standard $\Lambda$CDM paradigm, the classical picture of matter accumulation on halo-scales involves matter collapsing under its own gravity to form DM overdensities (``haloes''), which are the formation site for stars and galaxies. The eventual result is a filamentary distribution of haloes in a structure referred to as the ``cosmic web''. Hierarchical mass assembly of haloes and galaxies is one of defining processes predicted by a $\Lambda$CDM cosmology, with numerical work from \citet{Genel2010} using the Millennium and Millennium-II simulations suggesting that $60$\% of total halo mass growth is the result of mergers (in a DM-only regime). With an implementation of baryonic physics, \citet{Voort2011a} show that gas accretion onto haloes is predominantly ``smooth-mode'' (that is, not from resolved mergers), with merger-driven baryon growth (with mass ratios above $1:10$) being only significant in groups and clusters.


\subsection{Quantifying baryon accretion: theoretical and observational literature}
\label{sec:intro:literature}
The balance between gas inflows and outflows define the ``baryon cycle'' - the time evolution and breakdown of baryons (and their properties) in different reservoirs, at both galaxy- and halo-scales. The baryon cycle eventually sets the scaling between the stellar mass, gas mass and halo mass of galaxies, with recent simulation-based work highlighting the need for a more complete understanding of the interaction between inflows and outflows in the circum-galactic medium (CGM, see \citealt{Mitchell2019} and \citealt{Mitchell2020}). 
Continuity arguments form the basis of many phenomenological efforts to model the baryon cycle in a cosmological context \citep{Bouche2010,Dave2012,Lilly2013,Peng2014}. Analytical approaches typically involve a halo inflow term, and the eventual accretion onto the central galaxy is counteracted by preventative feedback terms. As an illustration, \citet{Dave2012} adopt the following prescription:
\begin{equation}
  \dot{M}_{\rm in}= \dot{M}_{\rm grav}-\dot{M}_{\rm prev}+\dot{M}_{\rm recyc},
  \label{eq:1}
\end{equation}
where $\dot{M}_{\rm in}$ is the net inflow rate to the galaxy, $\dot{M}_{\rm grav}$ is the total mass inflow rate onto the host halo (see Equation \ref{eq:2}), $\dot{M}_{\rm prev}$ represents the growth rate of intra-halo/CGM gas (gas inside the halo, but outside the central galaxy), and $\dot{M}_{\rm recyc}$ is the rate at which previously ejected gas (now part of the CGM) is reincorporated into the central galaxy's inter-stellar medium (ISM). The preventative feedback parameter, $\zeta$, is then defined to be $\zeta\equiv1-\frac{\dot{M}_{\rm prev}}{\dot{M_{\rm grav}}}$. The $\dot{M}_{\rm grav}$ term is derived from Extended Press Schechter (EPS) theory \citep{Neistein2006}, taking into account the assumed cosmology and predicted clustering of matter into spherical haloes. It is described by Equation \ref{eq:2} from \citet{Dekel2009}:
\begin{equation}
  \frac{\dot{M}_{\rm grav}}{M_{\rm halo}}=0.47 f_{\rm b}\left(\frac{M_{\rm halo}}{10^{12}M_{\odot}}\right) ^{0.15}\left(\frac{1+z}{3}\right)^{2.25}\ {\rm Gyr^{-1}}.
  \label{eq:2}
\end{equation}

\noindent We note that this formulation is dependent on $\sigma_{8}$ and $\Omega_{\rm m}$ cosmological parameters, which in this case were taken from WMAP5 \citep{WMAP5}. Since these values are very similar to the new Planck concordance parameters \citep{Planck2015,Planck2018}, which themselves have small uncertainties ($\approx1$\% and $\approx5$\% on $\sigma_{8}$ and $\Omega_{\rm m}$ respectively), it is unlikely that the behaviour of this formulation would change significantly with the choice of cosmology. The functional form of Equation \ref{eq:2} highlights the following points: (i) the efficiency of gas inflow $\left(\dot{M}_{\rm grav}/{M_{\rm halo}}\right)$ onto haloes (predicted from DM-only arguments) increases modestly with halo mass, (ii) the efficiency of gas inflow onto haloes decreases over cosmic time, and (iii) the amount of gas that enters a galaxy's ISM is modulated by preventative feedback from the galaxy. The latter can be broken down into: photoionisation, stellar feedback (winds \& heating), AGN feedback (winds \& heating) and thermal pressure from gravitational compression. More recently, some empirical models have begun to allow a connection between galaxy star formation rates (SFRs) and halo-scale inflow rates \citep{Moster2018,Behroozi2019}. 

Another approach to model halo mass growth rates is presented in \citet{Correa2015c} with the COMMAH code. This approach is also based on EPS theory \citep{Correa2015a,Neistein2006}, but also folds in the requirement for a defined relation between halo concentration and formation time. This relation was obtained through the analysis and fitting of results from a collection of hydrodynamical and DM-only simulations with varying cosmological parameters \citep{Correa2015b}. \citet{Correa2015c} found small changes in results when cosmological parameters were varied, 
with tension at maximum $\approx 0.1$~dex for a given redshift. 

Several simulation-based studies have investigated the influence of feedback physics on gas accretion. It is important to differentiate between gas accretion onto {\it haloes} and gas accretion onto {\it galaxies} (effectively, the $\dot{M}_{\rm grav}$ term and $\dot{M}_{\rm in}$ term in the \citealt{Dave2012} prescription, respectively). The interplay between feedback mechanisms and gas inflow within the halo environment is explored in depth in \citet{Voort2011a}, who focus on central {\it halo}- and {\it galaxy}-scale mass inflow (gross mass flux of gas particles into these structures over a discrete timestep). They define haloes as per the \subfind\ algorithm \citep{Springel2001,Dolag2009}, and the ISM as gas that is dense enough (most of which will be star forming) within $0.15 R_{\rm 200}$. With a range of \gadget-based \citep{Springel2005} hydrodynamical simulations implementing varying physics, they find: (i) that the efficiency of gas accretion onto {\it haloes} is only weakly dependent on halo mass (and at halo masses above $10^{11}M_{\rm halo}$, is fairly insensitive to feedback), and (ii) that accretion rates to {\it galaxies} are much more sensitive to somewhat uncertain feedback/cooling mechanisms.

\citet{Nelson2015} use two simulations based on the moving mesh {\sc arepo} code, one with both stellar \& AGN feedback (similar to the {\sc ILLUSTRIS} simulation), and one with neither stellar nor AGN feedback. They focus on inflows at {\it galaxy}-scales for centrals, and measure {\it net} accretion rate using a tracer-particle method, counting the number of inward crossings of a boundary at $r=0.15r_{\rm vir}$, subtract the number of outgoing crossings. They find that the fraction of mass delivered via smooth accretion is consistently lower in the presence of feedback by a factor of $2$ at all redshifts, and that for haloes of mass $10^{11.3}M_{\odot}<M_{\rm halo}<10^{11.4}M_{\odot}$ that inflow rates to the virial sphere are similar between their feedback and no-feedback runs at early times, but diverge by $\approx 0.25$~dex at $z\approx 0$, where inflow rates are suppressed in the presence of feedback. 

More recently, \citet{Correa2018} investigate gas accretion onto haloes and central galaxies in the \eagle\ suite of hydrodynamical simulations \citep{Schaye2015,Crain2015}. To calculate accretion rates to each halo, they find the flux of particles from outside the Friends-Of-Friends (FOF) halo to within the virial sphere, $R_{\rm 200,\ crit}$. With this methodology, they find that gas accretion at the halo-scale in the reference physics run deviates from a DM-only prediction \citep{Correa2015a,Correa2015b,Correa2015c}, with inflow suppressed for lower mass haloes. Interestingly, they show at the {\it galaxy}-scale for $z<2$ for haloes with  $10^{11.7}M_{\odot}<M_{\rm halo}<10^{12.7}M_{\odot}$ that the accretion rate to galaxies remains roughly independent of halo mass, while for haloes of mass $M_{\rm halo}<10^{11.7}M_{\odot}$ or $M_{\rm halo}>10^{12.7}M_{\odot}$, there is an obvious positive scaling of accretion rate with halo mass. They attribute the flattening at the galaxy-scale to AGN feedback, as this flattening vanishes in the no-AGN variant. They do not focus on comparing accretion rates onto haloes between different runs with varying feedback physics. 

Focusing instead on satellite galaxies, gas accretion and its environmental dependence at the galaxy-scale was investigated in \citet{Voort2017} using the reference \eagle\ simulation \citep{Schaye2015,Crain2015}. Using adjacent snapshots, they define gas accretion onto the galaxy to be the number of newly star-forming ISM particles found in a satellite, which were not star-forming at the previous snapshot. They found galaxies which become satellites in massive haloes can be ``starved'' of gas accretion, directly causing their star formation to quench.

\citet{AnglesAlcazar2017} used the high-resolution FIRE simulations to explore in-detail the baryon cycle of a small sample of galaxies, in particular the contribution of different accretion channels (e.g. ``fresh'' accretion, ``inter-galactic transfer'', and ``merger'' channels) over a wide range of galaxy mass. They find that the contribution of main progenitor recycling typically increases towards late times, together with the inter-galactic transfer component. \citet{Hafen2019}, using the FIRE-2 simulations, show that inter-galactic medium (IGM) accretion and winds from central galaxies typically dominate the origin of gas in the CGM, with increasing importance of satellite wind and stripping at higher halo masses. \citet{Hafen2020} subsequently show that while most CGM gas either remains in the CGM or is accreted to the central galaxy, a significant CGM gas fraction can be accreted to satellite galaxies (up to $3-4\%$ at $M_{\rm halo}\approx10^{12}M_{\odot}$). 

On a more statistical basis, the breakdown of gas accretion channels was explored in \citet{Mitchell2020} using the reference $100$~Mpc \eagle\ simulation run at both the halo- and galaxy-scale. They also show that the dominant contributor to mass accretion are first-infall and recycled particles, with the importance of mergers greatest for higher mass haloes, $M_{\rm halo}\gtrsim10^{12}M_{\odot}$. A similar simulation-based analysis of DM accretion channels, and the influence baryonic physics on baryon accretion channels, is currently lacking in the literature.

Although simulation-based methodologies of measuring inflow are decidedly diverse, it is also clear that the efficiency of inflow scales with the mass of a halo in a non-linear manner, eventually acting to set the relationship between stellar and halo mass. Most simulation-based work has indicated that feedback has little impact on the accretion rates at the {\it halo}-scale, unlike the {\it galaxy}-scale. The reader should note that the technique of calculating inflow, binning, and averaging can have significant impact on the quantitative measurements of accretion rates, and that some quantitative tension between works could plausibly be explained by the specific methodology used.

Observational constraints on gas accretion onto haloes and galaxies are limited. Direct measurement of inflow is exceedingly difficult given the typically small covering fraction of accreting material, the predominance of outflows, and the necessity for individual high signal-to-noise (S/N) spectra \citep{Rubin2012}. Additionally, accreting matter typically has a relatively weak kinematic signature compared to the galaxy's systematic redshift, unlike (typically high-velocity) feedback-driven outflows. Observational studies have confirmed the occurrence of inflows around the Milky Way (e.g. \citealt{LehnerHowk2011,Lockman2008}). Others have investigated extra-galactic sources with various methodologies, for instance direct ``down-the-barrel'' measurements of redshifted absorption lines (e.g. \citealt{Rubin2012}, for a review see \citealt{Rubin2017}), and quasar-galaxy pairs (e.g. \citealt{Bouche2013,Bouche2016,Ho2017,Rahmani2018,Zabl2019}). At this stage, studies of this nature are difficult to discuss quantitatively, with some work focusing on individual galaxies, and others limited to a sample of tens of galaxies \citep{Giguere2017}. Despite the scarcity of direct observations, it is clear from the star formation histories and rates of galaxies that accretion is expected to be a major regulator of galaxy growth \citep{Lilly2013}, and that accretion and outflows are likely interacting in the CGM (e.g. \citealt{Martin2019,Kacprzak2019,Pointon2019,Nielsen2020}).

In this study, we measured halo-scale accretion rates in an extensive selection of the simulation runs available from the \eagle\ suite, using the robust structure finder, \velociraptor\ \citep{Elahi2011,Elahi2019a}. We developed a self-consistent methodology to measure particle influx and the breakdown of this influx into physically distinct channels (``first-infall'', ``recycled'', ``transfer'', and merger'' components of inflow: see \S \ref{sec:methods:chumm}) between runs, which allowed us to systematically explore the influence of sub-grid physics on inflow rates to haloes and their origin, with a focus on comparing the behaviour of baryons and DM.

The content we present in this paper is structured as follows. In \S \ref{sec:methods}, we introduce (i) the \eagle\ hydrodynamical simulation suite and the sub-grid models that are relevant to this study, (ii) \velociraptor\ and \treefrog: the phase-space structure finder we use to identify bound haloes and substructures (and its accompanying halo merger tree generator), and (iii) \chumm: the code we use to calculate and analyse accretion rates onto haloes in \eagle. \S \ref{sec:s3} compares the build-up of gas and DM in \eagle\ runs with fiducial sub-grid physics. \S \ref{sec:s4} looks at the impact of including various physical processes in the hydrodynamical simulation when compared to the results of \S \ref{sec:s3}. In \S \ref{sec:discussion}, we discuss the implications of our findings on semi-analytic prescriptions of baryon build-up in galaxies. In \S \ref{sec:conclusions}, we conclude what these findings tell us about the growth of baryonic matter in galaxies, and future directions. Lastly, in Appendices \ref{sec:appendix:massfunctions}, \ref{sec:appendix:resolution}, \ref{sec:appendix:dt} and \ref{sec:appendix:r200}, we explore the sensitivity of our measurements to (i) number statistics, (ii) simulation spatial/mass resolution, (iii) simulation temporal resolution, and (iv) measurement methodology, respectively. 

\section{Methods}\label{sec:methods}
\subsection{The \eagle\ simulations}\label{sec:methods:eagle}
\begin{table*}
\begin{center}
\begin{tabular}{ |c||c|c|c|c|c|c|c|c|c|c|c| } 
 \hline
 {Run Name} &\shortstack{$ L_{\rm box}$\\${\rm [cMPc]}$} & \shortstack{ \\$N_{\rm part}$ } & \shortstack{$m_{\rm DM}$\\$[M_{\odot}]$} & \shortstack{$m_{\rm gas}$\\$[M_{\odot}]$}&
\shortstack{$\epsilon$\\$[{\rm pkpc}]$} &\shortstack{$f_{\rm th}$\\(min)}&\shortstack{$f_{\rm th}$\\(max)}&\shortstack{$\Delta T_{\rm SN}$\\$[{\rm K}]$}& \shortstack{$\Delta T_{\rm AGN}$\\$[{\rm K}]$} & RC & \shortstack{$N_{\rm field\ halo}\ (z=0)$\\($M_{\rm halo}>10^9M_{\odot}$)} \\ 
 \hline
 L25-REF & $25$ & $376^3$ & $9.7\times10^6$ & $1.8\times10^6$&$0.7$&0.3&3&$10^{7.5}$&$10^{8.5}$&$\checkmark$&$9,807$\\ 
 \hline
 L50-REF & $50$ & $752^3$ & $9.7\times10^6$ & $1.8\times10^6$&$0.7$&0.3&3&$10^{7.5}$&$10^{8.5}$&$\checkmark$&$74,743$\\ 
 \hline
 L50-NOAGN & $50$ & $752^3$ & $9.7\times10^6$ & $1.8\times10^6$&$0.7$&0.3&3&$10^{7.5}$&N/A&$\checkmark$&$74,351$\\ 
 \hline
 L50-AGNdT8 & $50$ & $752^3$ & $9.7\times10^6$ & $1.8\times10^6$&$0.7$&0.3&3&$10^{7.5}$&$10^{8}$&$\checkmark$&$74,788$\\ 
 \hline
 L50-AGNdT9 & $50$ & $752^3$ & $9.7\times10^6$ & $1.8\times10^6$&$0.7$&0.3&3&$10^{7.5}$&$10^{9}$&$\checkmark$&$75,645$\\ 
 \hline
 L25-ONLYAGN& $25$ & $376^3$ & $9.7\times10^6$ & $1.8\times10^6$&$0.7$&N/A&N/A&N/A&$10^{8.5}$&$\checkmark$&$9,944$\\ 
 \hline
 L25-WEAKSN & $25$ & $376^3$ & $9.7\times10^6$ & $1.8\times10^6$&$0.7$&0.15&1.5&$10^{7.5}$&$10^{8.5}$&$\checkmark$&$9,951$\\ 
 \hline
 L25-STRONGSN & $25$ & $376^3$ & $9.7\times10^6$ & $1.8\times10^6$&$0.7$&0.6&6&$10^{7.5}$&$10^{8.5}$&$\checkmark$&$9,839$\\ 
 \hline
 L25-NOFB & $25$ & $376^3$ & $9.7\times10^6$ & $1.8\times10^6$&$0.7$&N/A&N/A&N/A&N/A&$\checkmark$&$9,362$\\ 
 \hline
 L25-NONRAD & $25$ & $376^3$ & $9.7\times10^6$ & $1.8\times10^6$&$0.7$&N/A&N/A&N/A&N/A&$\times$&$10,751$\\ 
\hline

\end{tabular}
\end{center}
\caption{Simulation parameters for the \eagle\ runs utilised in this paper \protect{\citep{Schaye2015,Crain2015}}. $L_{\rm box}$ is the comoving box size of the simulation; $N_{\rm part}$ refers to the number of DM particles (and initial number of gas particles); $M_{\rm DM}$ and $M_{\rm gas}$ refer to the masses of DM and gas particles in the simulation respectively; $\epsilon$ refers to the Plummer equivalent maximum gravitational smoothing length; $f_{\rm th,\ min}$ \& $f_{\rm th,\ max}$ are the minimum and maximum supernovae energy transfer fractions (per unit stellar mass); $\Delta T_{\rm SN}$ \& $\Delta T_{\rm AGN}$ are the heating temperatures adopted for stellar and AGN feedback; ``RC'' indicates the inclusion of radiative cooling; and $N_{\rm field\ halo}\ (z=0)$ describes the number of field haloes in each run at the final snapshot.}
\label{tab:sims}
\end{table*}

\begin{table*}
\begin{center}
\begin{tabular}{ |c||c|c|c|c|c|c|c|c|c|c|c| } 
 \hline
 {Run Name} &\shortstack{$ L_{\rm box}$\\${\rm [cMPc]}$} & \shortstack{ \\$N_{\rm part}$ } & \shortstack{$m_{\rm DM}$\\$[M_{\odot}]$} & \shortstack{$m_{\rm gas}$\\$[M_{\odot}]$}&
\shortstack{$\epsilon$\\$[{\rm pkpc}]$} &\shortstack{$f_{\rm th}$\\(min)}&\shortstack{$f_{\rm th}$\\(max)}&\shortstack{$\Delta T_{\rm SN}$\\$[{\rm K}]$}& \shortstack{$\Delta T_{\rm AGN}$\\$[{\rm K}]$} & RC & \shortstack{$N_{\rm field\ halo}\ (z=0)$\\($M_{\rm halo}>10^9M_{\odot}$)} \\ 
 \hline
 L25N752-REF & $25$ & $752^3$ & $1.2\times10^6$ & $2.3\times10^5$&$0.35$&0.3&3&$10^{7.5}$&$10^{8.5}$&$\checkmark$&$9,906$\\ 
 \hline
 L25N752-RECAL  & $25$ & $752^3$ & $1.2\times10^6$ & $2.3\times10^5$&$0.35$&0.3&3&$10^{7.5}$&$10^{9}$&$\checkmark$&$10,068$\\ 
 \hline
  (\gadget)\ L32-NONRAD  & $32$ & $512^3$ & $2.7\times10^7$ & $5.0\times10^6$&$3.0$&N/A&N/A&N/A&N/A&$\times$&$74,724$\\ 
\hline

\end{tabular}
\end{center}
\caption{Simulation parameters for the additional convergence and testing runs utilised in this paper, with \eagle\ data from \protect{\citet{Schaye2015,Crain2015}}. $L_{\rm box}$ is the comoving box size of the simulation; $N_{\rm part}$ refers to the number of DM particles (and initial number of gas particles); $M_{\rm DM}$ and $M_{\rm gas}$ refer to the masses of DM and gas particles in the simulation respectively; $\epsilon$ refers to the Plummer equivalent maximum gravitational smoothing length, $f_{\rm th,\ min}$ \& $f_{\rm th,\ max}$ are the minimum and maximum supernovae energy transfer fractions (per unit stellar mass); $\Delta T_{\rm SN}$ \& $\Delta T_{\rm AGN}$ are the heating temperatures adopted for stellar and AGN feedback; ``RC'' indicates the inclusion of radiative cooling; and $N_{\rm field\ halo}\ (z=0)$ describes the number of field haloes in each run at the final snapshot. Note that the last simulation, L32-NONRAD, is not an \eagle\ run, rather a locally run \gadget-based 32 Mpc adiabatic physics box.}
\label{tab:sims2}
\end{table*}

 The \eagle\ (Evolution and Assembly of GaLaxies and their Environments) simulation suite \citep{Schaye2015,Crain2015} is a collection of cosmological hydrodynamical simulations that follow the evolution of galaxies and cosmological structure down to $z=0$.  The ANARCHY \citep{Schaller2015} set of revisions (designed to correct for ``classical'' SPH issues) were implemented on the \gadget-3 tree-SPH code \citep{Springel2005} to perform the \eagle\ simulations over a variety of periodic volumes and resolutions. \eagle\ adopts the parameters of a ${\Lambda}$CDM universe from \citet{Planck2014}, with initial conditions outlined in \citet{Jenkins2013}. Sub-grid physics modules were implemented to treat processes that are important for galaxy formation and evolution, but occur on scales below the resolution-scale of the simulation. These include: (i) radiative cooling and photoheating, (ii) star formation, (iii) stellar evolution and enrichment, (iv) stellar feedback, and (v) supermassive black hole (SMBH) growth and AGN feedback. Below, we provide a brief description of how these mechanisms are modelled in \eagle. 

Photo-heating and radiative cooling are applied based on the work of \citet{Wiersma2009}. This includes the effect of 11 elements in H, He, C, N, O, Ne, Mg, Si, S, Ca, and Fe \citep{Schaye2015}. The effect of radiation from the UV and X-ray background described by \citet{Haardt2001} is implemented on each element individually. Since the \eagle\ simulations do not provide the resolution to model cold, interstellar gas, a density-dependent temperature floor (normalised to $T=8,000$~K at $n_{\rm H}=10^{-1}{\rm cm}^{-3}$) is imposed. 

To model star formation, a metallicity-dependent density threshold is set, above which, star formation is locally permitted \citep{Schaye2015}. Gas particles are converted to star particles stochastically, with the star formation rate based on a tuned pressure law \citep{Schaye2008}, calibrated to the work of \citet{Kennicutt1998} at $z = 0$. The energy feedback from star formation is treated with a thermal energy injection of $10^{51}$ erg per type Ia supernovae (SNIa) event, the amount of which is a function of the IMF adopted \citep{Chabrier2003}. This is implemented in the form of a temperature boost to the surrounding particles of  ${\Delta}T_{\rm SF} = 10^{7.5}K$, based on the work of \citet{DallaVecchia2012}. The number of stars heated is calculated using Equation \ref{eq:sfb}, taken from Equation 8 in \citet{Schaye2015}:
 
\begin{equation}
  \langle N_{\rm heat}\rangle \approx 1.3f_{\rm th}\left( \frac{\Delta T}{10^{7.5}\rm\ K}  \right)^{-1},
  \label{eq:sfb}
\end{equation}

\noindent{where $f_{\rm th}$ is the fraction of the total amount of energy from core collapse supernovae per unit stellar mass that is injected on average. $f_{\rm th}$ varies between set minimum and maximum values (see Tables \ref{tab:sims} and \ref{tab:sims2}), the value in this range calculated based on local ISM properties. }

 SMBHs are seeded in \eagle\ when a DM halo exceeds a virial mass of $10^{10}\,\rm h^{-1} M_{\odot}$, with the seed SMBHs having an initial mass of $10^{5}\,\rm h^{-1} M_{\odot}$. Subsequently, SMBHs can grow via Eddington-limited-accretion \citep{Schaye2015}, as well as mergers with other SMBHs, according to work by \citet{Springel2005b}.  Similar to stellar feedback, AGN feedback in \eagle\ also involves the injection of thermal energy into particles surrounding the SMBH in the form of temperature boost of ${\Delta}T_{\rm BH}=10^{8.5}$K (in the reference physics  run; \citealt{Schaye2015}). The rate of energy injection (which directly sets the number of heated particles) from AGN is determined using the SMBH accretion rate, and a fixed energy conversion efficiency, as in Equation \ref{eq:agnfb}:
 
 \begin{equation}
\frac{\Delta E}{\Delta t}=\epsilon_{\rm f}\epsilon_{\rm r}\ \dot{m}_{\rm accr}c^{2},
  \label{eq:agnfb}
\end{equation}

\noindent{where $\dot{m}_{\rm accr}$ is a modified Bondi-Hoyle accretion rate (see Equations 9,10 in \citealt{Schaye2015}), and $\epsilon_{\rm f}\epsilon_{\rm r}=0.015$.}

\ \\ There are several free parameters in a number of the \eagle\ sub-grid modules, and these are calibrated (separately for the standard resolution runs and the L25N752-RECAL run) to match $z\approx0$ observations of (i) the galaxy stellar mass function, (ii) the galaxy size-mass relation, and (iii) the galaxy BH mass - stellar mass relation. We make use of 10 standard resolution \eagle\ runs, all with varying feedback physics (see Table \ref{tab:sims}) but identical starting mass resolution, and a number of supplementary runs with varying mass and temporal resolution (see Table \ref{tab:sims2}). For the remainder of the paper, we refer to each run by their identifier in the ``run-name'' columns for brevity. 

We include the mass distributions of field haloes in $4$ of the standard and supplementary runs in Appendix \ref{sec:appendix:massfunctions} - specifically those of L50-REF, L25-REF, L32-NONRAD, and L25N752-RECAL. Mass distributions for all standard resolution L50 (L25) runs are comparable to the L50 (L25)-REF distribution. We check the convergence of our results with the higher resolution \eagle\ variants, with $752^{3}$ particles in a $25$Mpc box in Appendix \ref{sec:appendix:resolution}. We also use a non-radiative \gadget-based $32$ Mpc box (with $512^3$ particles) to check temporal convergence (Appendix \ref{sec:appendix:dt}), and for enhanced number statistics compared to the \eagle\ L25-NONRAD box in \S \ref{sec:s4}. 

\subsection{Structure finding \& halo trees with \velociraptor\ and \treefrog}\label{sec:methods:velociraptortreefrog}
We identify haloes and subhaloes in the \eagle\ runs using \velociraptor\ \citep{Elahi2011,Elahi2019a}, a 6D friends of friends (6D-FOF) structure finding algorithm (that is to say, we do not make use of halo and subhalo information provided in the \eagle\ public catalogues, outlined in \citealt{McAlpine2016}). \velociraptor\ first uses a 3D-FOF algorithm \citep{Davis1985} to identify field haloes. Two particles $i$ and $j$ are ``linked'' together if they satisfy the condition: 
\begin{equation}
\frac{({\bf x}_{i}-{\bf x}_{j})^2}{l_{x}^{2}}<1
\ \ ({\rm where\ } l_{x} = 0.2 \times {\rm inter-particle\ spacing)}.
\label{eq:3}    
\end{equation} 
3D-FOF algorithms can spuriously link dynamically distinct objects together through tenuous particle bridges. Thus, for each 3D-FOF object, \velociraptor\ subsequently applies a 6D-FOF algorithm (including spatial and velocity information) in order to separate virialised structures \citep{Elahi2019a}. The velocity dispersion is calculated for each 3D-FOF object $k$, $\sigma_{v,\ k}$, and particles within the 6D-FOF are linked together if:
\begin{equation}
    \frac{({\bf x}_{i}-{\bf x}_{j})^2}{l_{x}^{2}}+\frac{({\bf v}_{i}-{\bf v}_{j})^2}{l_{v}^{2}}<1
    \ \ ({\rm where}\ l_v=\alpha_{v}\sigma_{v,\ k}\ {\rm and}\ \alpha_{v}\approx1.25).
    \label{eq:4}
\end{equation}
An unprocessed 3D-FOF halo will typically contain numerous density peaks, some of which may reside outside the virial radius centred on the inner-most minimum potential. Once the 6D-FOF algorithm has been run over a 3D-FOF object, each of the nested density peaks will be identified as ``sub-haloes'' of the parent halo (with the notable exception of the central 6D-FOF object, which remains identified as the parent halo).

To link haloes through time, we use the halo merger tree code \treefrog\ \citep{Elahi2019b}, developed to work on the outputs of \velociraptor. This code compares the particles in haloes across multiple snapshots by calculating a ``merit'' based on the fraction of particles that are shared by two (sub)haloes $i$ and $j$ at different times. In circumstances where several matches are identified for one (sub)halo with similar merits (e.g. mergers of several similar mass haloes), \treefrog\ ranks particles based on their binding energy. This combined information is then used to estimate a merit function, which makes use of the total number of particles shared, and their binding energy information (see Eq. 3 in \citealt{Elahi2019b}). 

\subsection{Accretion calculations with \chumm}\label{sec:methods:chumm}
In order to calculate accretion rates onto haloes in our simulations, we developed and used the python-based repository \chumm\ (Code for Halo AccUMulation of Mass, available at \url{https://github.com/RJWright25/CHUMM}). Methodologies to calculate inflow rates to structure in cosmological simulations can be categorised as either (i) Eulerian, or (ii) Lagrangian in nature. The Eulerian method involves an instantaneous calculation of inflowing mass flux at a defined boundary, while the Lagrangian approach involves a calculation of inflowing mass flux, over a discrete time interval. Our method for calculating accretion flux is Lagrangian in nature, with a number of (optional) additional conditions imposed on inflow particles. As discussed by \citet{Mitchell2019} in the context of outflows, the Lagrangian method yields more accurate measurement of time-integrated flux, and is less sensitive to the stochasticity of instantaneous particle behaviour at the boundary of haloes. 

With Lagrangian accretion approaches, it is important to consider the time interval used. In previous literature, \citet{Voort2011a} use intervals of up to $\approx 1.5\ {\rm Gyr}$, while \citet{Voort2017} and \citet{Correa2018} use a varying $\Delta t$, corresponding to adjacent \eagle\ snapshots. We also elect to calculate accretion over the interval between adjacent \eagle\ snapshots ($29$ snapshots from $z=20$ to $z=0$), corresponding to a $\Delta t$ ranging between $\approx 250\ {\rm Myr}$ at minimum (at $z\approx4$), and $\approx 0.9\ {\rm Gyr}$ at maximum (at $z\approx0$). The varying timesteps in \eagle\ that we use to calculate accretion rates  correspond to $0.5-1.0$ times a halo dynamical time across the full redshift range that we analyse, and thus, for the purposes of our work, we don't believe there is a need for higher cadence intervals. To investigate the sensitivity of our accretion calculations to the time interval, we used a higher cadence ($200$ snapshot) non-radiative simulation and looked at temporal convergence in Appendix \ref{sec:appendix:dt}. We find that while the normalisation of accretion rates can change slightly with the interval, our results are not sensitive to the choice. Additionally, we refer the reader to the work of \citet{Mitchell2020}, who analyse inflow with higher cadence full-physics \eagle\ runs and find similar temporal convergence properties to our L32-NONRAD run.

To identify haloes, \chumm\ either uses the outputs from \velociraptor\ and selects particle members of 6D-FOF objects, or \chumm\ can use a spherical overdensity (SO) criterion, where particles within a spherical region defined by $R_{\rm vir}$ are selected as constituting a halo. By default, \velociraptor\ returns particle lists for field haloes, with substructure particles removed or ``deblended''. For field haloes with substructure, we define the full FOF (with substructure) by re-introducing substructure particles to the host. For the SO criterion, we use the \velociraptor\ halo catalogue to iterate through each halo, $i$, determine its central position ${\bf x}_{i}$ and virial radius, $r_{200,\ i}$, and then search for the particles $j$ (with position ${\bf x}_{j}$) that satisfy the condition $\left(\frac{|{\bf x}_{i}-{\bf x}_{j}|}{r_{200,\ i}}\right)<f$, where \chumm\ can take a list of user defined $f$ values. Particles satisfying this condition then constitute the SO definition of the halo, $i$. 

We compare results using different halo definitions (i.e. 6D-FOF particles or spherical overdensity regions with boundaries at fractions of $R_{200}$) in Appendix \ref{sec:appendix:r200}, and find results to change slightly (but not qualitatively) between methods. We elect to proceed with the FOF-based boundary for accretion, as we believe this more consistently reflects the nature of accretion over different halo mass scales. 

 \begin{figure*}
\includegraphics[width=1\textwidth]{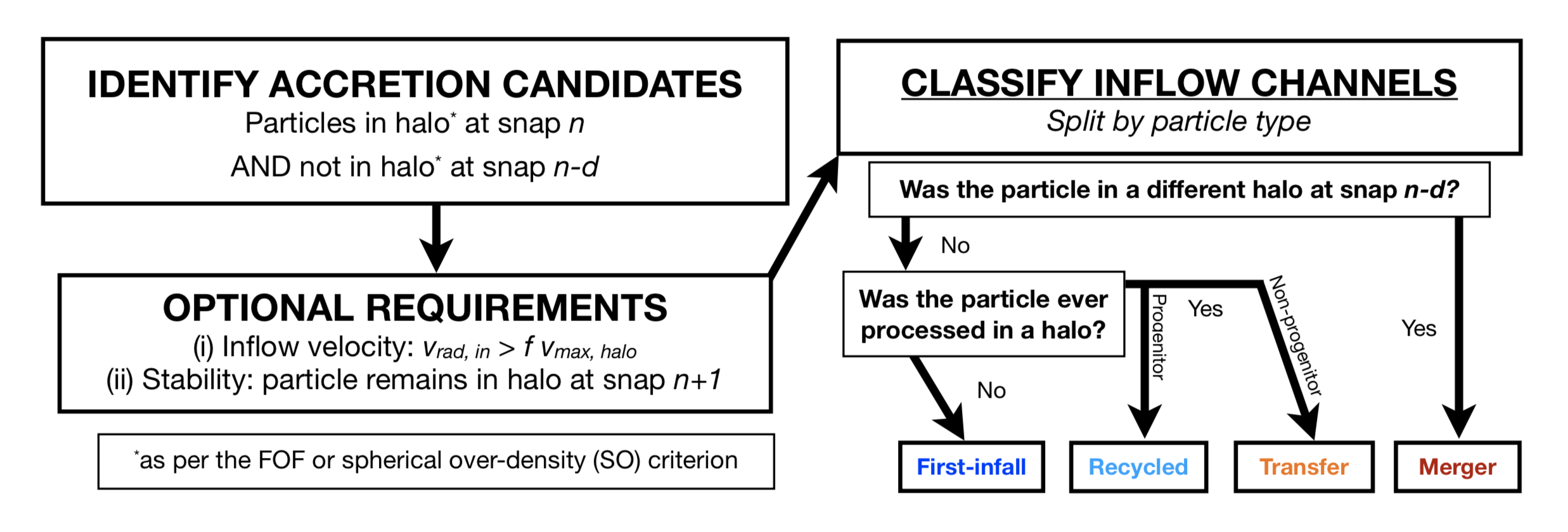}
\caption{A schematic of our algorithm \chumm\ to calculate accretion rates to haloes. We identify accretion candidates by finding particles which were not in a halo at a snap $n-d$, but joined the halo at snap $n$. We are able to impose a number of extra criteria on these particles before decomposing the total accretion rate into ``channels'' of accretion based on the processing history of each particle. The code is publicly available at \url{https://github.com/RJWright25/CHUMM}.}
\label{fig:methods:chumm}
\end{figure*}

The algorithm \chumm\ uses to calculate accretion rates is summarised in Figure \ref{fig:methods:chumm}. Recall each simulation has a finite number of snapshots $N$ (hereafter snaps) that encode the physical state of the system at times $t_{0}<t<t_{N}$, and we equate snap $n$ with the state of the system at lookback time $t=t_{n}$. To calculate accretion rates onto haloes at snap $n$, we identify accretion candidates as the particles that exist in the halo at snap $n$ as per the definition above, but did not exist in the halo at snap $n-d$ ($d$ being the snap ``depth'' of the calculation, i.e. the difference in snap index, chosen as 1 unless otherwise stated). The summed mass of these candidate particles, normalised by $\Delta t = t_{n-d} - t_{n}$ (where $t_{n}$ represents the lookback time at snap $n$), constitutes the raw {\it gross} total accretion rate of the halo at snap $n$. Accretion rates are split by particle type, with the particle type categorised at the initial snap $n-d$, before undergoing any processing in the halo (such that gas particles at snap $n-d$ which were transformed to star particles by snap $n$ would be considered gas inflow, not stellar inflow). 

We experimented with imposing a number of radial velocity cuts on accretion candidates at the initial snap $n- d$ (to ensure particles are travelling towards the halo's center-of-mass at a sufficient rate): $v_{\rm rad,\ in,\ n-d}>f\times v_{\rm max,\ halo}$. We found that for a range of reasonable $f$ fractions of the maximum halo circular velocity $v_{\rm max,\ halo}$, our inflow results were not significantly impacted. We also experimented with requiring accretion candidates to remain in the halo they were considered accreted to at the subsequent snapshot (i.e. to remain in the halo for an additional $t_{n}-t_{n+1}$ to be considered as ``stable'' accretion), and again saw only a slight change in the normalisation of accretion rates, rather than any qualitative alterations to trends. When testing this requirement, the effect was quantitatively the largest ($\approx 0.1$~dex) on the non-radiative boxes, where stochastic flux of particles at the halo boundary (both inwards and outwards) are more common-place in the absence of radiative cooling. Even in the non-radiative runs, however, we saw no qualitative change in the results. As such, we elected not impose the velocity or stability criteria for our results (unless otherwise stated), in order to make our results minimally complex to interpret. 

Using the FOF-based accretion algorithm, we are able to categorise the nature of the inflow particles (their accretion ``channel'') based on (i) their host at snap $n-d$, and (ii) their processing history. The particle's host at snap $n-d$ determines the origin of accretion as either from the field, or from another \velociraptor\ structure (the particles of each origin we refer to as {\it ``cosmological''/``smooth''} or {\it``merger''/``clumpy''} accretion particles respectively). 

For the ``cosmological''/``smooth'' accretion case, a particle is considered {\it ``processed''} if it has existed in any halo (as defined by \velociraptor) up to and including snap $n-d$ (the initial snap), and {\it ``unprocessed''/``first infall''} otherwise. Commonly the term ``pristine'' is used to describe the accretion channel of metal-poor, un-enriched particles, however we elect to use the term ``first infall'' for our ``unprocessed'' channel to recognise that these particles are only ``unprocessed'' insofar as \velociraptor's ability to identify bound structures, ultimately limited by the finite mass resolution of the simulation. We show the relevant peak in the halo mass functions from each run from \velociraptor\ in Appendix \ref{sec:appendix:massfunctions}, corresponding to $M_{\rm halo}\approx 10^{8.5}M_{\odot}$ at $z\approx0$ in the standard resolution \eagle\ variants.

We can further decompose the ``processed'' channel of cosmological accretion into a ``recycled'' and ``transfer'' component - for particles which were previously processed in a progenitor (main, non-main, or a satellite) halo ({\it ``recycled''} accretion) and those that were previously processed in an unrelated halo ({\it ``transferred''} accretion) respectively. In a future paper, we explore the physical properties of accreting particles as a function of inflow channels, and find that processed particles (according to our definition) are clearly metal-enriched relative to the first-infall particles, giving us confidence that our methodology is physically-motivated.

While our definitions of accretion channels are similar to those in \citep{Mitchell2020}, we remark that the definition of ``transfer'' in previous literature can be different, for example in \citet{AnglesAlcazar2017}. In their work, ``transfer'' is considered as all accretion of particles that have been processed in haloes other than the main progenitor of halo $i$ by snapshot $n$. This therefore includes the accretion of particles that were processed in (secondary) progenitor halos that merge into $i$. We think it makes more physical sense to group all halos that end up merging by snapshot $n$ and consider particles processed in those as part of ``recycling'' rather than ``transfer''. This distinction is quite significant, as the fraction of mass that is processed in non-main progenitors is of the same order of that contributed by the main progenitor.

\section{Baryon and dark matter build-up in haloes with fiducial physics}\label{sec:s3}
In this section we focus on analysing the accretion rate of baryons and DM onto haloes in the \eagle\ simulation suite, in particular the fiducial physics implementation in L50-REF. We approach this topic with the aim of understanding the differential manner in which baryons and DM are accreted onto haloes, and their main channels of accretion. In \S \ref{sec:s4}, we explore the same problem, focusing instead on changes which can arise based on the feedback models implemented in the simulation. We remark that in this paper, we do not focus on the physical properties of accreted matter (for instance their temperature, density and metallicity), and defer a full investigation of these properties in the context of halo inflow for a future paper. 

\subsection{Comparing our algorithm and past literature}

\begin{figure*}
\includegraphics[width=1\textwidth]{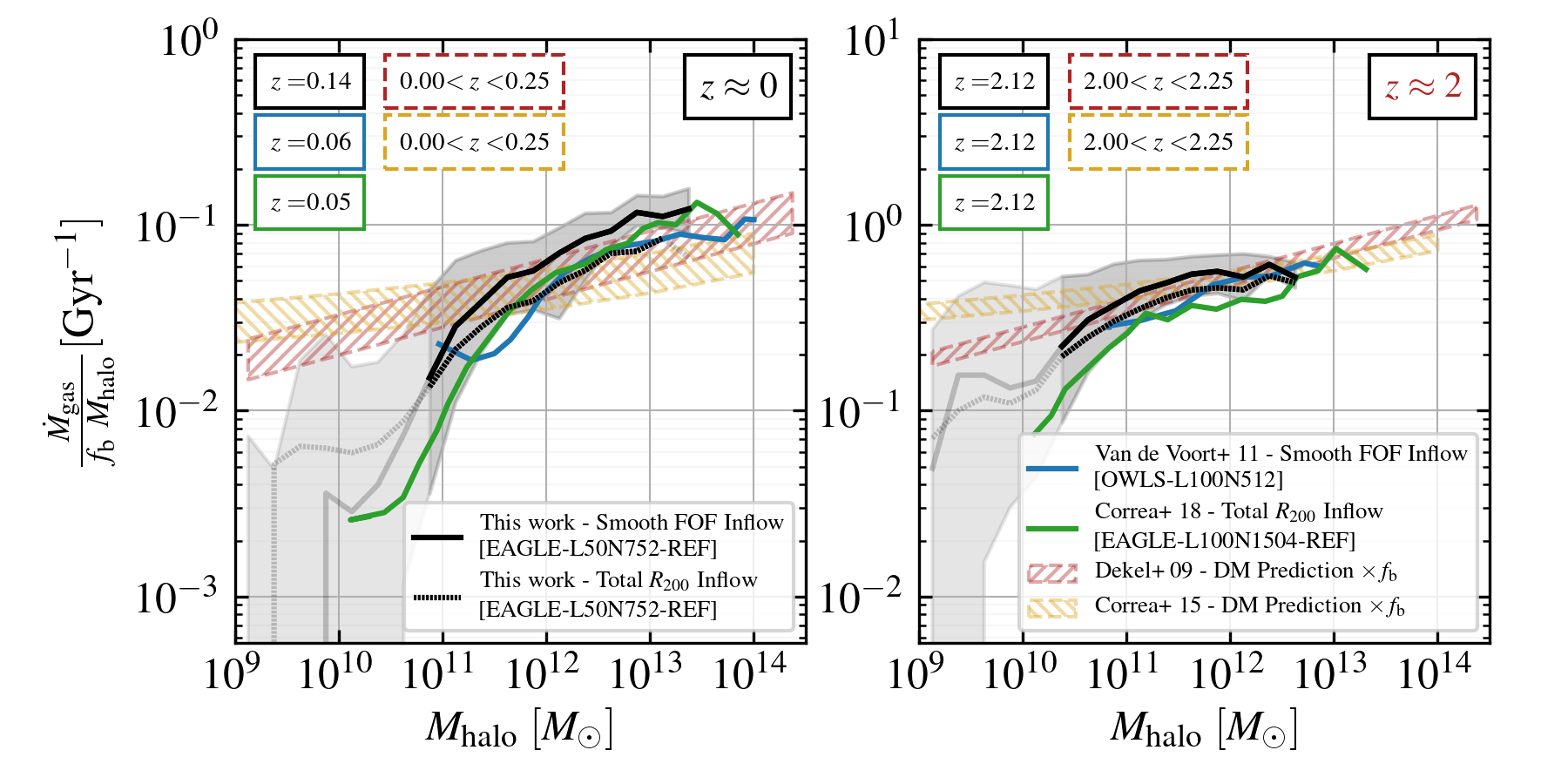}
\caption{Smooth baryon inflow efficiencies from our algorithm (FOF, solid black lines; and $R_{200}$, densely dotted black lines) for field haloes at $z\approx0$ (left) and $z\approx2$ (right), compared to inflow rates from \citet{Voort2011a}, \citet{Correa2018}, and those predicted based on DM inflow rates from \protect{\citet{Dekel2009}} (right-slanting hatched regions) and \protect{\citet{Correa2015a,Correa2015b,Correa2015c}} (left slanting hatched regions). We note that each panel has slightly altered axes limits. Grey shaded regions correspond to the $16^{\rm th}-84^{\rm th}$ percentile ranges associated with our L50-REF FOF accretion rate calculation. Line transparency has been increased where the average efficiency has been calculated from a bin in which more than $50$\% of haloes were subject to an accretion flux of less than 50 gas particles. We use $24$ evenly log-spaced bins of halo mass from $10^{9}M_{\odot}$ - $10^{15}M_{\odot}$, and take the median accretion rate in each of these bins, requiring at least $5$ objects to take this average. Our smooth inflow rate refers exclusively to particles {\it not} accreted from a different FOF structure. Results from \protect{\citet{Voort2011a}} include only smooth accretion, however data from \protect{\citet{Correa2018}} include the merger channel of accretion. The predictions from \protect{\citet{Dekel2009}} and \protect{\citet{Correa2015a,Correa2015b,Correa2015c}}  (both shown hatched regions, indicating predictions within the given redshift range) are analytic prescriptions of halo mass growth scaled by $f_{\rm b}$ - and as such, also includes merger-based/clumpy accretion. In general, our measurements are qualitatively consistent with previous works, and quantitatively within $0.5$~dex for well resolved halo mass bins. }

\label{fig:s3:literaturecomparison1}
\end{figure*}

 The results to follow, unless otherwise stated, focus on field haloes in the standard resolution \eagle\ runs, as outlined in Table \ref{tab:sims}. For an investigation of numerical convergence with higher resolution, we direct the reader to Appendix~\ref{sec:appendix:resolution}. Our primary method of measuring accretion rates uses the FOF inflow algorithm ({\it including} accretion onto satellite subhaloes), however we direct the reader to Appendix {\ref{sec:appendix:r200}} for an overview of how using a spherical-overdensity inflow calculation (or, only including accretion onto the central subhalo with the FOF algorithm) would influence our results. Additionally, we attempt to minimise the cuts we impose to our accretion calculations in order to make the results easier to interpret. As such, we stress that in the work to follow, we do not impose any velocity cuts on our results, nor do we impose the stability requirement. 
 
 We also remark that when we refer to the halo mass of field haloes (associated with FOF inflow), we use the total mass of all particles identified as part of the FOF, {\it not} just those within $R_{\rm 200,\ crit}$ (with the exception of halo mass in the context of our spherical-overdensity accretion algorithm, for which we use $M_{\rm 200,\ crit}$). For FOF masses below $10^{13}M_{\odot}$, $M_{\rm 200,\ crit}$ corresponds very tightly with $M_{\rm FOF}$, but for FOF masses of $>10^{13}M_{\odot}$, $M_{\rm 200,\ crit}$ is systematically lower than $M_{\rm FOF}$ by up to $50\%$. The method of binning and averaging has significant bearing on the numerical results we present: we indicate the bins we use in the caption of each figure, and only take averages in bins where there are $\geq5$ objects. For information on the number statistics we use over halo mass, we show the halo mass distributions for relevant runs in Appendix \ref{sec:appendix:massfunctions}. Lastly, where possible, we show transparent lines where accretion rates have been calculated from a flux of less than $50$ particles. 

We first compare our results using the fiducial physics L50-REF box to the previous work of \citet{Voort2011a}, \citet{Correa2018} and the DM-based predictions from \citet{Dekel2009} (using WMAP5 cosmological parameters, \citealt{WMAP5}) and \citet{Correa2015a,Correa2015b,Correa2015c}  (using 2015 Planck cosmological parameters, \citealt{Planck2015}) in Figure \ref{fig:s3:literaturecomparison1}. As discussed in \S \ref{sec:intro:literature}, the techniques adopted in the past to measure inflow in simulations (and the simulations themselves) are diverse, and it is important to understand the influence of the methodology on inflow measurements. Figure \ref{fig:s3:literaturecomparison1} shows gas accretion efficiency (gas accretion rate normalised by halo mass and cosmological $f_{\rm b}$) as a function of halo mass, for redshifts $z\approx0$ and $z\approx2$. Our results using the FOF-based algorithm (solid black lines) include only the smooth component accretion (accretion from particles not identified as part of a \velociraptor\ FOF structure at the previous snapshot), while our results using the $R_{200}$-based algorithm (dotted lines) include particles of all origins. The redshift we quote for our calculations is the mean redshift of the two snaps used to calculate accretion over, and where possible, we quote the redshift of previous works in the same way. As outlined in \S \ref{sec:intro:literature}, previously used methods of calculating inflow rates in simulations are diverse - \citet{Voort2011a} use a FOF halo-based method of identifying inflow particles in the {\sc OWLS} simulations, while \citet{Correa2018} calculate require mass flux from particles which start outside the FOF and end up inside the virial radius, $R_{\rm 200,\ crit}$, in the \eagle\ simulations. The results from \citet{Correa2018} include particles of all origins, while the results from \citet{Voort2011a} include only smooth inflow. We also illustrate the analytic mass-growth predictions from \citealt{Dekel2009} and \citet{Correa2015a,Correa2015b,Correa2015c} in Figure \ref{fig:s3:literaturecomparison1} with dashed lines, corresponding to the total expected halo mass growth scaled by a factor of $f_{\rm b}\approx 0.16$.

Our results using both the FOF and $R_{200}$ algorithms appear qualitatively consistent with the works of \citet{Voort2011a} and \citet{Correa2018}, with a trend of increasing accretion efficiency with halo mass and redshift. In agreement with these data, compared to the DM-based predictions from \citet{Dekel2009} and \citet{Correa2015a,Correa2015b,Correa2015c}, we observe significant reduction in gas accretion rates to less massive haloes, below $M_{\rm halo}\approx10^{11.5}M_{\odot}$. We see that the FOF algorithm predicts higher accretion rates than the $R_{200}$ method by a steady $\approx 0.15$~dex, for $M_{\rm halo}\gtrsim10^{11.5}M_{\odot}$. We attribute the higher accretion rates in our FOF algorithm compared to the SO method to be principally due to the inclusion of accretion onto satellites subhaloes associated with each of the field haloes. We investigate the difference between accretion onto the central subhalo (excluding satellites) and our primary FOF algorithm (including satellites) in Appendix \ref{sec:appendix:r200}, and see that accretion rates to the central subhalo are reduced compared to the primary method by $\approx0.1$~dex across much of the halo mass range. For the remainder of the paper, we elect to continue to include the accretion onto satellites as our primary method, since we are interested with accretion onto the entire FOF group.

Quantitatively, we see that our FOF method appears to predict slightly higher accretion rates at $z\approx0$ and $z\approx2$ by $0.1-0.2$~dex compared to \citet{Voort2011a} and \citet{Correa2018} (even when the latter results include merger-based accretion). We attribute the difference in our results compared to \citet{Correa2018} to the above discussion of including satellite accretion (as discussed above) - and if we only account for accretion onto the central subhaloes, the (already slight) tension with \citet{Correa2018} is greatly reduced. Unsurprisingly, when we use an $R_{200}$ spherical overdensity method (densely dotted lines), we see very good agreement with \citet{Correa2018}. Since \citet{Voort2011a} also use a FOF algorithm, the inclusion of satellite accretion does not explain our increased gas accretion rates compared with their results. \citet{Voort2011a} do, however, show that using a higher resolution simulation (closer to our primary \eagle\ L50-REF box, with a 50 Mpc box with $2\times512^3$ particles, instead of their primary 100 Mpc box with $2\times512^3$ particles), that gas accretion rates are enhanced by $\approx 0.1$~dex. Accretion rates in this higher resolution run show less tension with our fiducial \eagle\ results, but since the calibration of sub-grid models are directly dependent on resolution, it is non-trivial to determine the reason for this change. Any further tension between results are potentially attributable to (i) \velociraptor\ vs. \subfind\ halo catalogues, and (ii) for the $z\approx0$ panel, the slight differences in measurement redshifts. We also note that we have compared measurements with the work of \citet{Mitchell2020}, and find consistent results over redshift and halo mass.

The comparison presented here gives us confidence that our algorithm is giving results that are largely consistent with previous literature, and can therefore be used for an in-depth analysis comparing the rates and channels of baryon and DM accretion onto haloes.

\subsection{Comparing baryon and dark matter accretion}\label{sec:s3:baryondm}
In this section, we directly compare halo-scale baryonic and DM inflow rates in the \eagle\ L50-REF run, showing the influence that the baryon richness of accreting matter can have on resultant halo properties and inflow channels. 
\subsubsection{Influence on halo properties}
\label{sec:s3:baryondm:haloprops}
\begin{figure}
\includegraphics[width=0.45\textwidth]{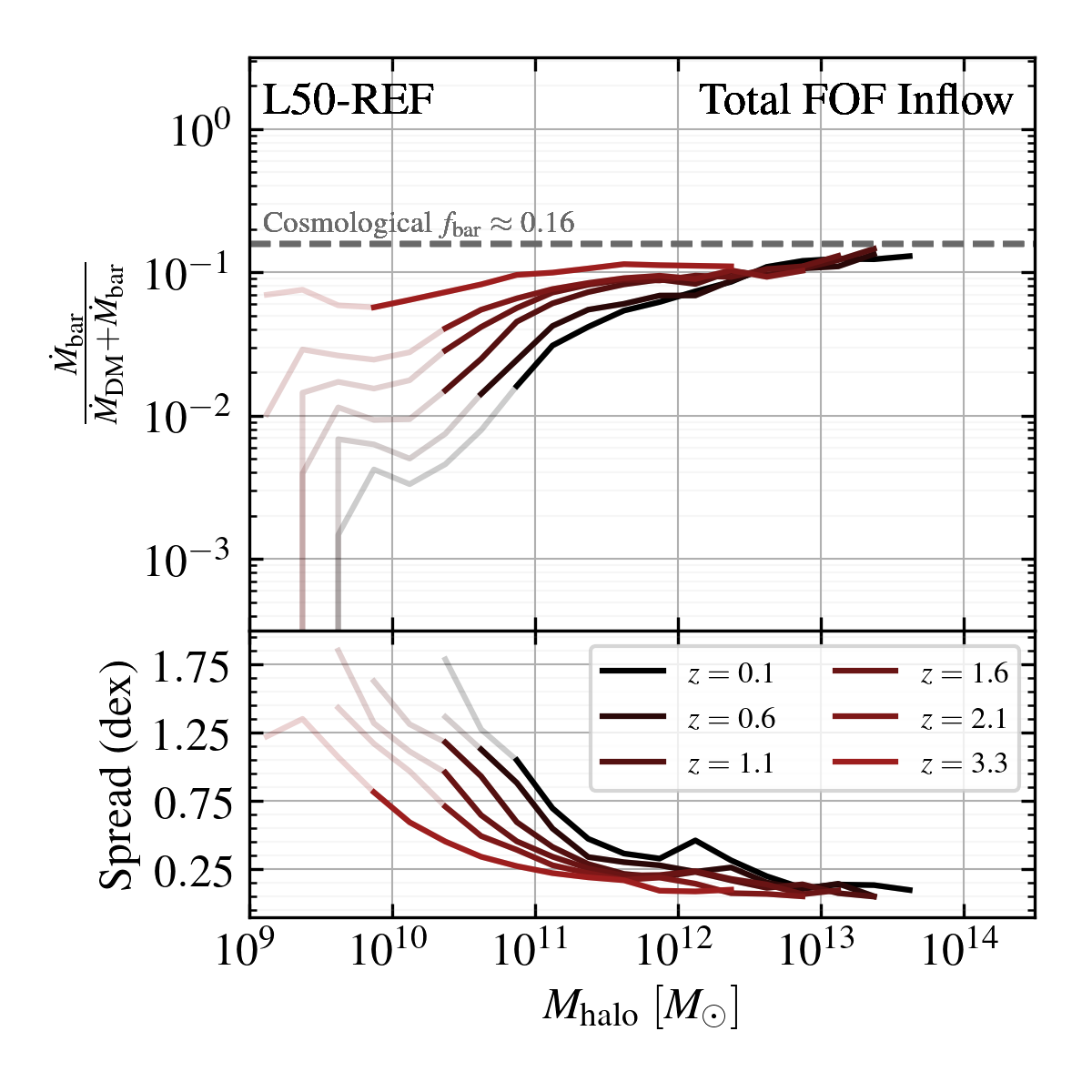}
\caption{Top panel: the baryon fraction of all halo-accreted matter (inflow baryon fraction, or $f_{\rm b,\ inflow}$) as a function of halo mass and redshift. Bottom panel: the $16^{\rm th}-84^{\rm th}$ percentile range of inflow baryon fraction in dex, as a function of halo mass and redshift. We use $24$ evenly log-spaced bins of halo mass from $10^{9}M_{\odot}$ - $10^{15}M_{\odot}$, and take the median accretion rate in each of these bins, requiring at least $5$ objects to take this average. Line transparency has been increased where the average efficiency has been calculated from a bin in which more than $50$\% of haloes were subject to an accretion flux of less than 50 gas particles. For each redshift selected, we see a trend of increasing $f_{\rm b,\ inflow}$ with halo mass up to $M_{\rm halo}\approx10^{12}M_{\odot}$, above which the median fraction plateaus slightly shy of the simulation cosmological baryon fraction. Inflow baryon fractions also universally increase with redshift, the dependence most notable for lower mass haloes, $M_{\rm halo}\lesssim10^{11.5}M_{\odot}$. The spread in $f_{\rm b,\ inflow}$ is largest in the same halo mass range, particularly at low redshift.}
\label{fig:s3:m200inflowfb}
\end{figure}

Above, we discussed the behaviour of gas accretion alone with halo mass and redshift. To directly compare the build-up of baryons and DM in haloes, we investigate the baryon fraction of matter accreted to field haloes in the \eagle\ L50-REF run as a function of halo mass and redshift in Figure \ref{fig:s3:m200inflowfb}. We calculate the baryon content of accreted matter ($f_{\rm b,\ inflow}$) for each halo by calculating the summed mass of accreted baryons, $\Delta M_{\rm bar,\ in}$, and normalising by the summed mass of all accreted material (DM and baryons),

\begin{equation}
    f_{\rm b,\ inflow}=\frac{\Delta M_{\rm bar,\ in}}{\Delta M_{\rm bar,\ in}+\Delta M_{\rm DM,\ in}}.
\label{eq:5}
\end{equation}

In the top panel of Figure \ref{fig:s3:m200inflowfb}, we show $f_{\rm b,\ inflow}$ against halo mass, using total inflow onto field haloes (including both smooth-mode inflow and merger-mode inflow). The baryon inflow rates which may be trivially expected based on DM inflow rates would correspond to $f_{\rm b,\ inflow}=f_{\rm b,\ cosmological}$ (grey dashed line). We find that for low mass haloes ($M_{\rm halo}\lesssim10^{12}M_{\odot}$), the composition of inflow material is baryon poor compared to the universal $f_{\rm b}$ - a finding consistent with the low gas accretion efficiencies in this mass range first presented in \citet{Voort2011a} (with OWLS) and \citet{Correa2018} (with reference \eagle\ physics). At $z\approx0$, the median baryon fraction of inflow at $M_{\rm halo}\approx 10^{11}M_{\odot}$ is only $\approx2$\%, roughly $10$ times lower than the universal baryon fraction $f_{\rm b}\approx 0.16$, and the halo baryon fractions predicted in \citet{Crain2007}. The baryon content of accreted matter increases with halo mass to $f_{\rm b}$ only for the most massive group sized haloes in the simulation, at $M_{\rm halo}\gtrsim 10^{13}M_{\odot}$. In addition, the baryon content of accreted matter increases with redshift for low mass haloes, $M_{\rm halo}\lesssim10^{11.5}M_{\odot}$, being roughly 1~dex higher at $M_{\rm halo}\approx10^{10.5}M_{\odot}$ at $z\approx2$ compared to $z\approx0$. Conversely, for higher mass haloes,  $M_{\rm halo}\gtrsim 10^{11.5}M_{\odot}$, there is little redshift dependence. In the bottom panel of Figure~\ref{fig:s3:m200inflowfb}, we show the spread (in dex) of the $16^{\rm th}-84^{\rm th}$ percentiles in $f_{\rm b,\ inflow}$. The spread in $f_{\rm b,\ inflow}$ is largest for low halo masses - at $z\approx0$, for $M_{\rm halo}\approx10^{11}M_{\odot}$, we see a spread of $1$~dex, while for $M_{\rm halo}\approx10^{13}M_{\odot}$ we see a spread of $0.2$~dex. This spread also decreases (for all halo masses) with increasing redshift. We find that the result of suppressed gas inflow (and larger variation in gas inflow) at low halo masses remains clear when we use the higher resolution \eagle\ L25N752-RECAL run (see Appendix \ref{sec:appendix:resolution}), indicating that this result is not a consequence of numerical effects. 

This result clearly begs the question: what physics in the simulation drives the depletion of halo baryon inflow, particularly for low mass haloes, $M_{\rm halo}\lesssim10^{11.5}M_{\odot}$? We investigate the origin of this effect (in the context of feedback) in \S \ref{sec:s4}, where we use various \eagle\ runs to investigate the effect on accretion rates in the presence and absence of these different feedback mechanisms. We find that in modern simulations, stellar and AGN feedback mechanisms are strong enough to influence gas dynamics at the halo-scale and modulate inflow rates. 

\begin{figure}
\includegraphics[width=0.45\textwidth]{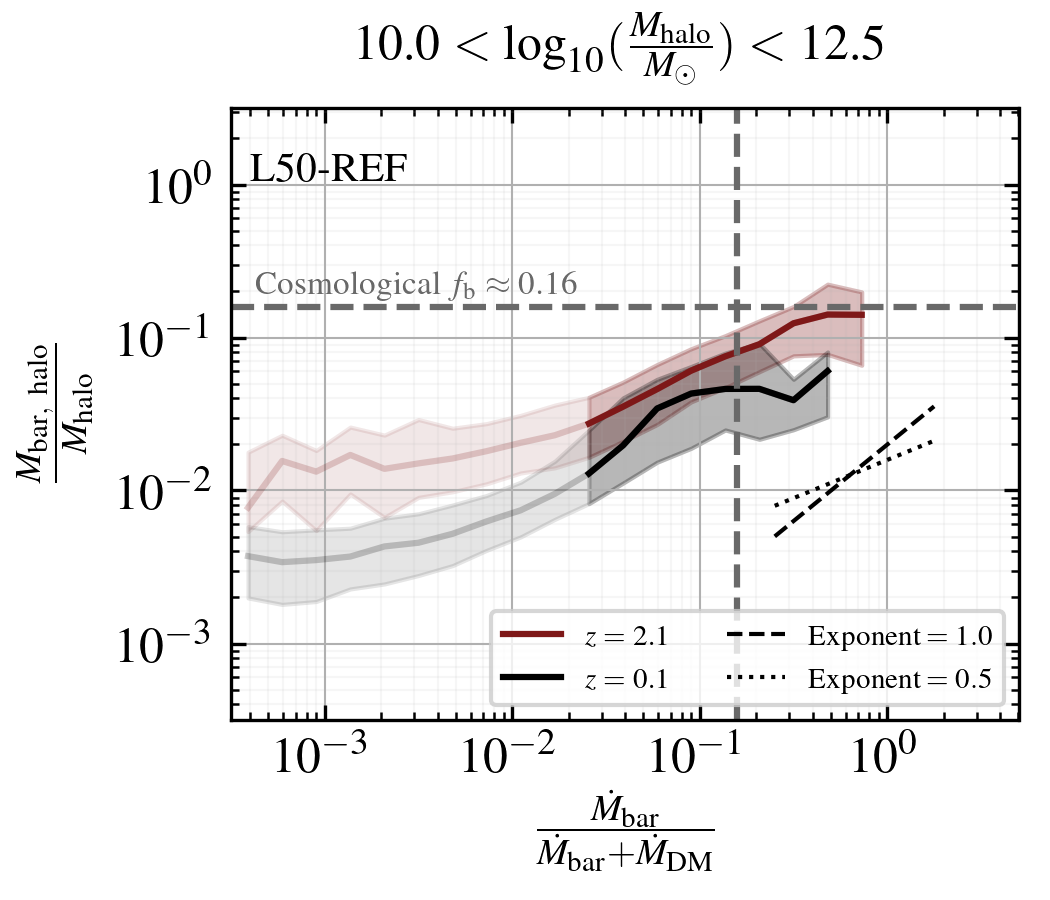}
\caption{The median halo-baryon fraction (including gas, stellar, and black hole particles in the FOF - see Equation~\ref{eq:6}) as a function of halo-inflow baryon fraction (see Equation~\ref{eq:5}), for field haloes in the mass range $10^{10.5}M_{\odot}<M_{\rm halo}<10^{12.0}M_{\odot}$ at $z\approx2$ and $z\approx0$. Line transparency has been increased where the average efficiency has been calculated from a bin in which more than $50$\% of haloes were subject to an accretion flux of less than 50 particles. Shaded regions correspond to the  $16^{\rm th}-84^{\rm th}$ percentile ranges. At both redshifts we observe a tight correlation between the accretion baryon fraction and the halo baryon fraction with $\lesssim0.5$~dex spread, the correlation becoming slightly steeper at $z\approx0$. The segments at the right of the panel show for reference a linear dependence and a power law of exponent $0.5$. At both redshifts, the slope is shallower than linear, and closer to a power law with an exponent of $0.5$.}
\label{fig:s3:inflowfbhalofb}
\end{figure}

\begin{figure*}
\includegraphics[width=0.89\textwidth]{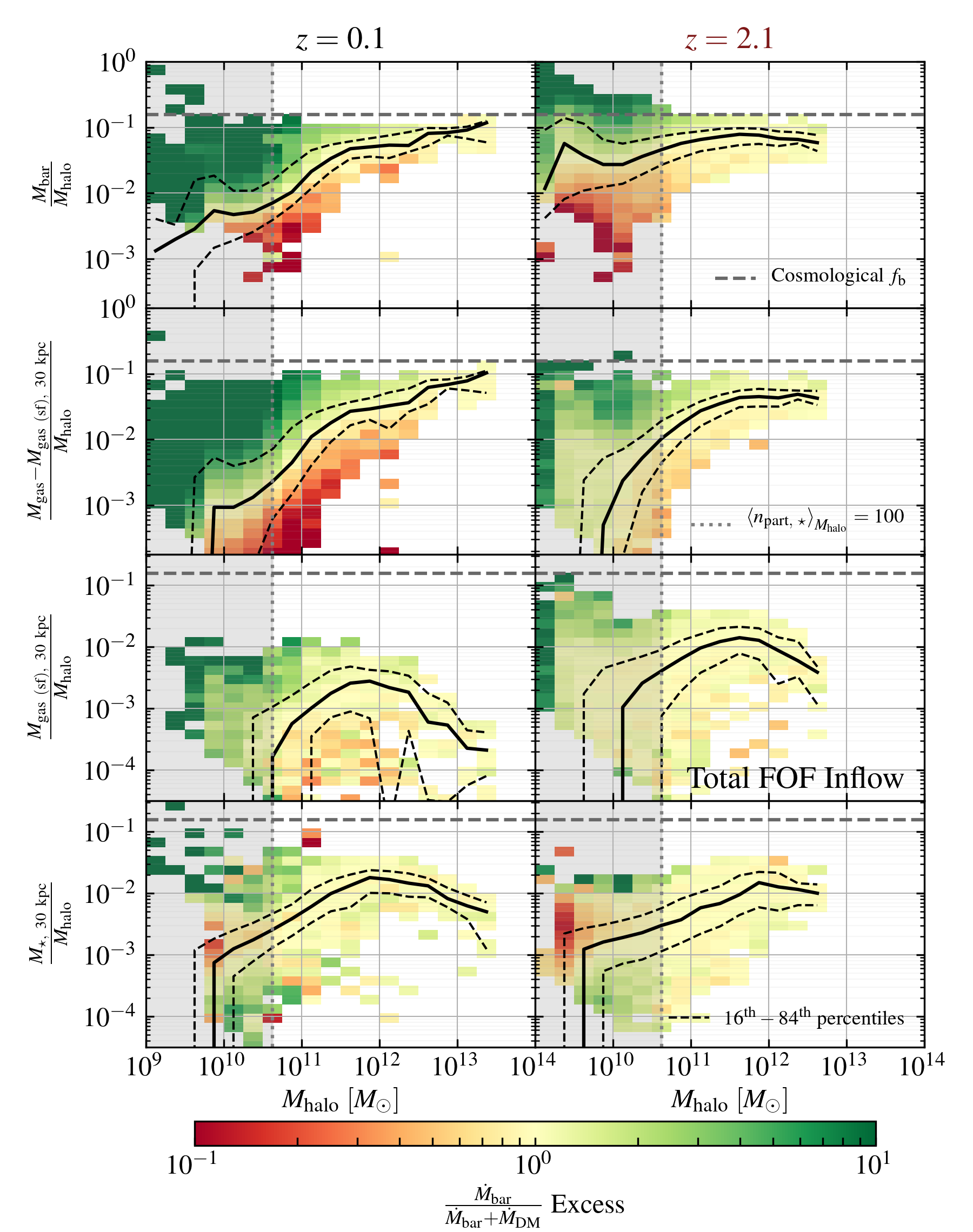}
\caption{The median fractions of mass contained in various baryon reservoirs, as a function of halo mass for field haloes (left panels: $z\approx0$, right panels: $z\approx2$). Dashed lines correspond to $16^{\rm th}-84^{\rm th}$ percentiles, and the grey shaded regions represents the mass range corresponding to haloes with, on average, less than $100$ star particles. We use $24$ evenly log-spaced bins of halo mass from $10^{9}M_{\odot}$ - $10^{15}M_{\odot}$, and take the median accretion rate in each of these bins, requiring at least $5$ objects to take this average. All panels are coloured by the median {\it excess} $f_{\rm b,\ inflow}$ in each halo mass bin. Top panels: the median total baryon fraction of field haloes as a function of halo mass. Second row: The median mass fraction in halo CGM gas (which we define as gas that is not star-forming in the central $30$ kpc of a halo). Third row: the median mass fraction in halo ISM gas (star-forming gas within the central $30$ kpc of a halo). Bottom panels: the median mass fraction in stars within the central $30$ kpc of a halo. We observe that for a fixed halo mass, the spread in total halo baryon fractions is well correlated with the baryon fraction of halo-accreted matter, with baryon depleted (rich) haloes typically being subject to baryon depleted (rich) inflow. We can see this correlation is strongest for the gas outside $30$ kpc and for non-star-forming (either warm or insufficiently dense) gas in the central $30$ kpc. However, the central star-forming gas and stellar component of the halo do not see significant enhancement as a result of instantaneous baryon inflow. }
\label{fig:s3:m200halofb}
\end{figure*}

To illustrate the impact that variations in baryon inflow has on the resulting halo population, in Figure~\ref{fig:s3:inflowfbhalofb} we show the baryon fraction of haloes (as defined in Equation \ref{eq:6}) as a function of their baryon inflow content, $f_{\rm b,\ inflow}$, at both $z\approx0$ and $z\approx2$. We restrict the mass range to $10^{10.0}M_{\odot}<M_{\rm halo}<10^{12.5}M_{\odot}$, where we see the greatest dynamic range in inflow baryon fractions as per Figure~\ref{fig:s3:m200inflowfb}, still with adequate resolution.

\begin{equation}
    f_{\rm b,\ halo}=\frac{[M_{\rm gas}+M_{\rm \star}+M_{\rm BH}]_{\rm \ FOF}}{M_{\rm tot,\ FOF}}.
\label{eq:6}
\end{equation}

 \noindent At both redshifts, we observe a tight correlation between $f_{\rm b,\ inflow}$ and $f_{\rm b,\ halo}$, with a roughly constant spread of $\approx 0.3-0.4$~dex (a factor of $2-3$) across the range of $f_{\rm b,\ inflow}$. The correlation is not linear, and appears to scale more closely with an exponent of $0.5$, i.e. $f_{\rm b,\ halo}\propto f_{\rm b,\ inflow}^{0.5}$ for both redshifts. While the slope is similar  for both redshifts, we see that at $z\approx2$, the dynamic range in $f_{\rm b,\ inflow}$ is narrower (as per Figure \ref{fig:s3:m200inflowfb}), and that $f_{\rm b,\ halo}$ is systematically higher. If we use a different mass cut and focus on higher mass haloes, $M_{\rm halo}\gtrsim10^{12}M_{\odot}$, the correlation remains (with a similar slope) - but over a much smaller dynamic range, from just below to the cosmological $f_{\rm b}$ for both $f_{\rm b,\ inflow}$ and $f_{\rm b,\ halo}$. This tight correlation indicates that the baryon-richness of a halo is clearly associated with baryon inflow onto that halo - and that {\it if we see variation in baryon inflow at the halo-scale, we would also expect to see this reflected in the baryon content and consequent properties of that halo.} The correlation in Figure~\ref{fig:s3:inflowfbhalofb} suggests that haloes may be baryon depleted {\it because} the material that gets accreted is baryon poor, and is  not exclusively a result of baryon processes internal to haloes and galaxies acting to remove the gas.  The latter definitely plays a role in modulating halo baryon content (see \citealt{Davies2019,Oppenheimer2020}), and will also influence the baryon content of infalling haloes (and consequent baryon inflow rates to their new host host), but the picture is clearly multi-faceted, with outflows and inflows quite possibly causally linked. Below, we further analyse the connection between the baryon content of haloes with the baryon content of the inflowing matter.

To investigate the distribution of baryons in field haloes and the consequence of the established variations in baryon inflow, in Figure~\ref{fig:s3:m200halofb} we show the median mass fractions of baryon particles in different intra-halo reservoirs (each row), as a function of their halo mass at $z\approx0$ (left) and $z\approx2$ (right). The parameter space in each panel is coloured by the median {\it excess} (calculated per halo mass bin) $f_{\rm b,\ inflow}$  in that parameter range - that is, how baryon rich the inflow is, compared to the median baryon fraction of the inflow at a given halo mass. As a function of halo mass, for each row of panels from top to bottom in Figure~\ref{fig:s3:m200halofb}, we show the mass fractions in: (i) all baryons, (ii) CGM gas (that is, all halo gas excluding star forming gas within the central $30$~kpc of the halo), (ii) star-forming ISM gas (star-forming gas within the central $30$ kpc of the halo center-of-mass), and (iv) stellar mass (within the central $30$ kpc of the halo center-of-mass). Each of these mass fractions are normalised by the total mass of the halo, $M_{\rm FOF}$. 

Focusing on the top row of panels in Figure~\ref{fig:s3:m200halofb}, we display the results seen in Figure \ref{fig:s3:inflowfbhalofb} as a function of halo mass. We observe that the spread in total halo baryon fraction at fixed halo mass (up to $M_{\rm halo}\approx10^{12}M_{\odot}$) is very well-correlated with the baryon fraction of accreted matter both at $z\approx0$ and $z\approx2$. Specifically, on average, baryon-depleted haloes appear to be host to baryon-depleted inflow, and baryon-rich haloes appear to be host to baryon-rich inflow. At higher halo masses, $M_{\rm halo}\gtrsim10^{12}M_{\odot}$, there is far less spread in halo baryon fractions, and most haloes are relatively baryon rich (approaching the cosmological $f_{\rm b}$ in the grey dashed line). The remaining bottom 3 rows of panels in Figure \ref{fig:s3:m200halofb} show where the baryonic matter is distributed in haloes, and also whether instantaneous baryon inflow has impact on the magnitude of these various reservoirs. If we observe a gradient in colour along the spread in mass fraction for a fixed halo mass, it indicates that baryon inflow plays a role in regulating this reservoir over the timescale when accretion was measured. 

Panels in the second row of in Figure~\ref{fig:s3:m200halofb} show the gas fraction in the CGM as a function of halo mass. The CGM gas fraction increases with halo mass, and is the primary baryonic constituent of haloes for those with masses above $M_{\rm halo}\approx10^{11}M_{\odot}$. We see that for the majority of the halo mass range (below $M_{\rm halo}\approx 10^{12.5}M_{\odot}$), the spread in CGM mass fraction at fixed mass is strongly correlated with $f_{\rm b,\ inflow}$). This is an unsurprising result: one would expect the majority of accreting baryons reside in the CGM at the snapshot subsequent to the accretion measurement. At a halo mass of $M_{\rm halo}\approx10^{12}M_{\odot}$, we see approximately $0.5$~dex of spread in between the $16^{\rm th}$ and $84^{\rm th}$ percentiles. The spread in $f_{\rm b,\ inflow}$ associated with these percentiles is also approximately $0.5$~dex at $z\approx0$ - telling us that variations in the baryon fraction of the inflowing matter could play a role in driving CGM gas content in reasonably massive haloes. 

The work of \citet{Davies2019} and \citet{Oppenheimer2020} indicate that variations in black hole mass and activity in \eagle\ are associated with variations in CGM gas fractions for Milky-Way like haloes (in the mass range $10^{12}M_{\odot}\lesssim M_{\rm halo}\lesssim10^{12.5}M_{\odot}$) due to AGN-driven {\it outflows}. In this mass bracket, we show that baryon {\it inflow} rates also vary strongly over the range of $f_{\rm CGM}$ values, which show a $16^{\rm th}-84^{\rm th}$ $f_{\rm CGM}$ percentile spread of $\approx0.5$~dex. We remark that when we produce the same figure instead using the L50-NOAGN run, we still see a correlation between CGM content and baryon inflow, however covering a smaller range in halo baryon fraction ($\approx 0.25$~dex), and slightly less variation in baryon inflow content. Based on these findings, together with those of \citet{Davies2019} and \citet{Oppenheimer2020}, it seems that ejective and preventative feedback work together continuously to modulate the baryon content of haloes, with feedback-driven winds possibly influencing infalling matter in this mass range.

In the third row \& fourth row of panels in Figure~\ref{fig:s3:m200halofb}, we show the mass fraction of each halo held in star-forming gas and stellar mass respectively, within the central $30$~kpc of a halo. This, to first order, corresponds to the cool/dense phase of the ISM, and the stellar component of the central galaxy, respectively. Over the well-resolved halo mass range, we observe that the spread in the mass fraction contributed by these mass reservoirs is not correlated with $f_{\rm b,\ inflow}$. This combination of findings tells us that while the baryon content of inflow has significant bearing on the instantaneous gas content of the CGM and potentially the warmer phase of central galaxy, over the timescale that we measure accretion ($\approx 1$~Gyr at $z=0$), the central galaxy's dense ISM and stellar mass content are not immediately influenced by how baryon rich the inflow is.

\subsubsection{Channels of baryonic and DM inflow}
\label{sec:s3:baryondm:channels}
\begin{figure*}
\includegraphics[width=0.92\textwidth]{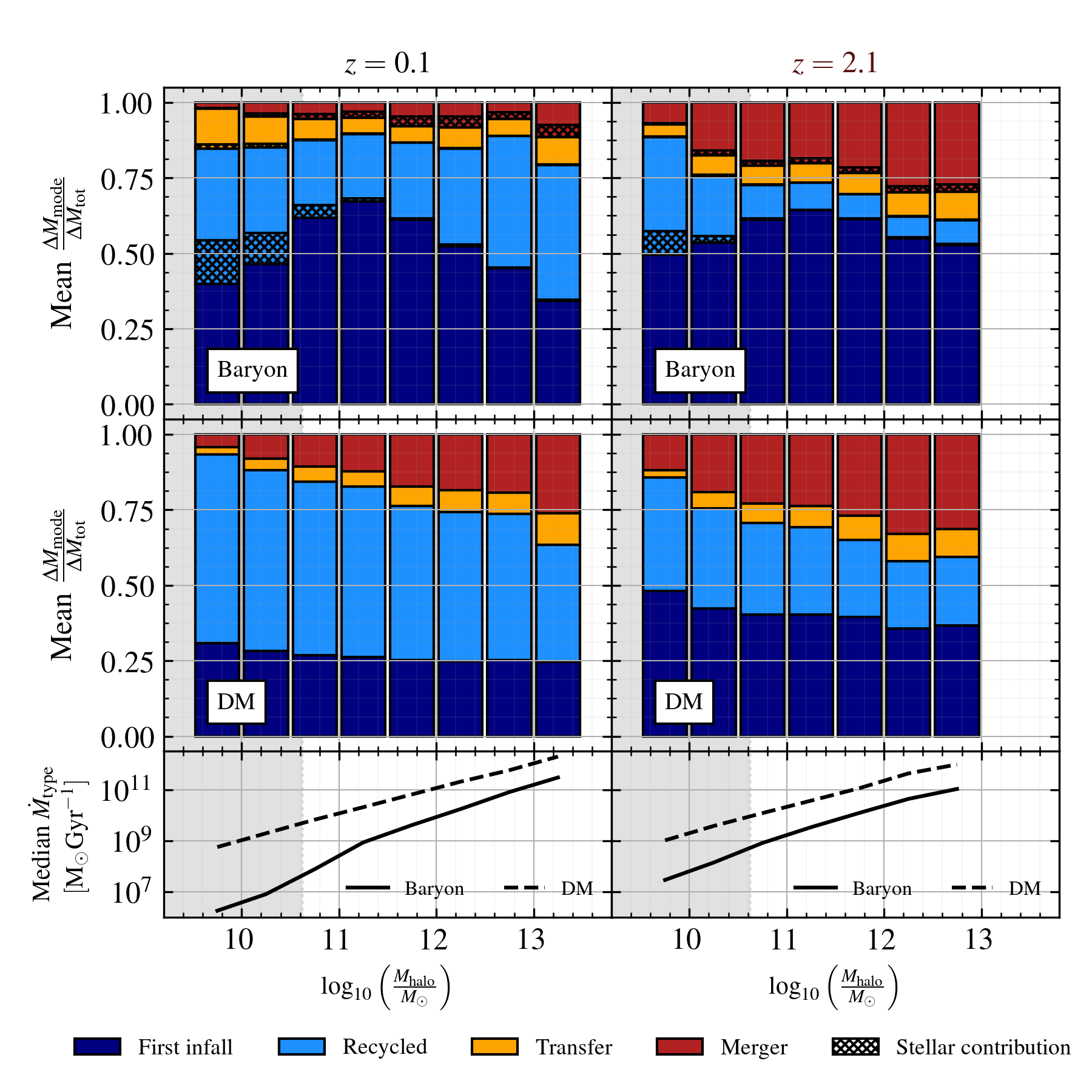}
\caption{Breakdown of the accretion channels of baryons (first row of panels) and DM (2nd row of panels) for $z\approx0$ (left panels), and $z\approx2$ (right panels) in the L50-REF run. The results are presented in $8$ equally log-spaced $M_{\rm halo}$ bins from $10^{9.5}M_{\odot}$ to $10^{13.5} M_{\odot}$, and we indicate the halo mass range in which each halo contains less than 100 stellar particles (where numerical issues could influence results) with grey shading. In the bottom panels we also illustrate the median raw accretion rate for each particle type in the same halo mass bins. Accretion channels are categorised as (i) first infall (completely unprocessed particles - blue), (ii) recycled (particles which have channels processed in some progenitor halo in the past, but were most recently accreted via smooth-mode cosmological accretion - green),  (iii) transfer (particles which have been processed in some non-progenitor halo in the past, but were most recently accreted via smooth-mode cosmological accretion - yellow) , or (iv) (particles which were, at the previous snap, part of another halo - red). We also illustrate the contribution of stellar accretion onto total baryon accretion with black hatched regions in the top panels. The sum of channels (i), (ii) and (iii) corresponds to ``smooth-mode'' accretion - that is, from particles not in a structure at the previous snapshot. We show the component of accretion contributed by stellar particles with the cross-hatched regions. The difference between DM and baryon accretion channels is most prevalent at late times, with mergers being roughly twice as dominant for DM inflow compared to baryon inflow for all halo mass bins. The proportion of recycled cosmological accretion increases towards late times for both DM and baryonic matter, and at all redshifts we see that DM is much more likely to have been pre-processed prior to accretion compared to accreting baryons, which are more often found to be on first infall to a halo.}
\label{fig:s3:accretionbreakdown}
\end{figure*}

In Figure~\ref{fig:s3:accretionbreakdown}, we compare the inflow channels of baryonic matter (top) to DM (bottom) in across halo mass for redshifts $z\approx0$ and $z\approx2$. We split inflow into 4 categories, namely: (i) first infall accretion: that is, inflow particles which had never previously been part of a \velociraptor\ halo structure (blue); (ii) recycled accretion: particles which have been processed in some progenitor halo in the past, but were most recently accreted via smooth-mode cosmological accretion (green); (iii) transfer accretion: particles which have been processed in some non-progenitor halo in the past, but were most recently accreted via smooth-mode cosmological accretion (yellow); or (iv)  merger/clumpy accretion, which at the previous snap were part of another halo (red). The sum of channels (i), (ii) and (iii) corresponds to smooth or cosmological accretion -  particles which were not in a structure at the previous snapshot. We illustrate the component of baryon accretion contributed by stellar particles with cross-hatched regions. 

Firstly, regarding the nature of stellar accretion channels, we find that the majority of stellar accretion (in the well-resolved halo mass range) occurs via mergers to high mass haloes, $M_{\rm halo}\gtrsim10^{12}M_{\odot}$, though the contribution to lower mass haloes becomes more significant at late times. We remind the reader that we classify infall particles by their type at the snap {\it prior} to accretion - but find, on average, that only a very small proportion ($\lesssim0.1\%$) of accreting gas is transformed to star particles at the subsequent snapshot in the halo mass range we analyse. We see significant stellar recycling at the low halo mass end ($M_{\rm halo}\lesssim10^{11}M_{\odot}$), but given that this corresponds to the regime where haloes contain $\lesssim100$ stellar particles, we cannot say with certainty that this is not a resolution driven feature. 

In general, the breakdown of accretion into its merger and smooth/cosmological (first infall, recycled and transfer) channels is somewhat similar for baryons compared to DM, the notable exception being higher mass haloes at late times, where the merger-mode is significantly more dominant for DM than baryons. At $z\approx0$, for haloes in the mass range $10^{12.5}M_{\odot}<M_{\rm halo}<10^{13.5}M_{\odot}$, we observe baryon inflow onto haloes to be $\approx 7\%$ merger-mode, compared to DM inflow which accumulates $\approx20\%$ via merger mode. This tells us that baryon accretion via mergers appears suppressed at late times, comparing to the behaviour we would expect based solely on DM. The physical interpretation of this result is that the infalling haloes contributing to the merger-based mass growth are already gas poor. We have already seen in Figure~\ref{fig:s3:m200halofb} that low-mass field haloes are, on average, baryon poor - at $z=0$ containing $\approx1\%$ baryons for haloes of mass $M_{\rm halo}\approx10^{11}M_{\odot}$, a factor of $>10$ below the universal baryon fraction. Thus, if these baryon-depleted haloes undergo a merger, most of the mass contributed to the descendant will be DM, rather than baryons. If the merger-mass is contributed by infalling satellite subhaloes, such satellites are known to be stripped of their gas in \eagle\ via environmental ``pre-processing'' prior to infall (see \citealt{Bahe2015,Bahe2019}). Our findings agree well with the work of \citet{Voort2011a}, who found that gas accretion onto haloes is predominantly ``smooth-mode'', with merger-driven baryon growth significant only in groups and clusters. The contribution from mergers on the DM accretion rate is smaller than the value found in \citet{Genel2010} for the Millennium and Millennium-II Simulations, which they quantified at $\approx 60\%$ independent of halo mass. This is a factor of $\approx 1.5-4$ times higher than what we find in \eagle. This is not entirely surprising, as \citet{Genel2010} reached the fraction above by extrapolating the merger rate to large mass ratios, while here we only account for ``resolved'' halo mergers. It is therefore likely that some fraction of smooth DM accretion corresponds to ``unresolved'' halo mergers. We believe that some of these ``unresolved'' halo mergers would correspond to the ``recycled'' mode of accretion (e.g. if the halo was resolved at some point in the simulation), and some would correspond to ``first infall'' mode accretion if the halo was never resolved. 

We note that the similar breakdown between smooth and merger components of baryon/DM accretion (aside from high mass haloes) is somewhat misleading, and decomposing smooth-mode inflow into its first infall (blue), recycled (green) and transfer (yellow) components illustrates the underlying disparity: we see that {\it accreted DM is significantly more likely to have been pre-processed in haloes than baryons for all redshifts and halo mass bins}. The proportion of baryons provided by first infall accretion appears to peak for both redshifts at $M_{\rm halo}\approx10^{11}M_{\odot}$ where first infall particles contribute  $\approx 60-65\%$ of all baryon accretion, the peak being slightly more prominent at $z\approx0$. The contribution of first infall-mode to DM accretion rates is fairly flat with halo mass, at $\approx 25\%$ for $z\approx0$ and $\approx40\%$ for $z\approx2$ - both significantly reduced by a factor of $\approx 2$ compared to baryons. The reduced first infall component for DM is met with a marked increase in the recycled inflow component, showing that it is fairly common for DM to be accreted by a halo, ejected from the FOF boundary, and subsequently be re-accreted to a descendant halo. As discussed above, while the classification of inflow into these channels is resolution dependant, we argue that the disparity seen comparing recycled-mode mass growth of baryons and DM is significant. In a future paper, we show that baryon particles in these different channels are indeed characterised by clearly different physical properties (particularly metallicity) and hence, can confidently conclude that our classification of accretion channels are physical, with resolution playing a minor role. We refer the reader to the work of \citealt{Mitchell2020} for an analysis of the breakdown of accretion channels as a function of resolution. Relating to our previously measured gas accretion rates in Figures~\ref{fig:s3:literaturecomparison1} and \ref{fig:s3:m200inflowfb}, the picture emerging is that baryons are much less likely than DM to be accreted at all redshifts (and thus, universally, less baryons are classified as processed), particularly for haloes with mass $M_{\rm halo}\lesssim10^{11.5}M_{\odot}$. 

We see that the contribution of inter-halo transfer accretion across halo mass is small ($\lesssim5\%$) for baryons and DM at both redshifts in the well resolved mass range, in agreement with \citet{Mitchell2020}. The transferred component does appear to increase with halo mass for DM, but remains fairly steady for baryons across halo mass. The difference in the transfer component between low- and high-mass haloes is likely due to the fact that low-mass haloes have a shallow potential well that fails to attract DM particles that have left other haloes, and in practice can only re-accrete part of what was originally expelled from the same halo. Consequently, a fraction of the ejected particles from low mass haloes ($\lesssim10^{11}M_{\odot}$), are likely contributing to the ``inter-halo transfer'' accretion component of the more massive haloes, $>10^{11}M_{\odot}$, with deeper potential wells. A physical explanation for the presence of transferred particles is so called ``cosmic-web stripping'' (e.g. \citealt{Benitez2013,Benitez2017}), which does not require feedback physics to eject particles - thus being particularly applicable to DM transfer. In \S \ref{sec:s4:channels} we explore this further, and discuss the strong influence that model physics has on baryonic accretion channels.

\subsection{Baryon accretion onto satellite subhaloes}
\label{sec:s3:subhaloes}
\begin{figure}
\includegraphics[trim=5mm 6.5mm 5mm 15mm, clip,width=0.475\textwidth]{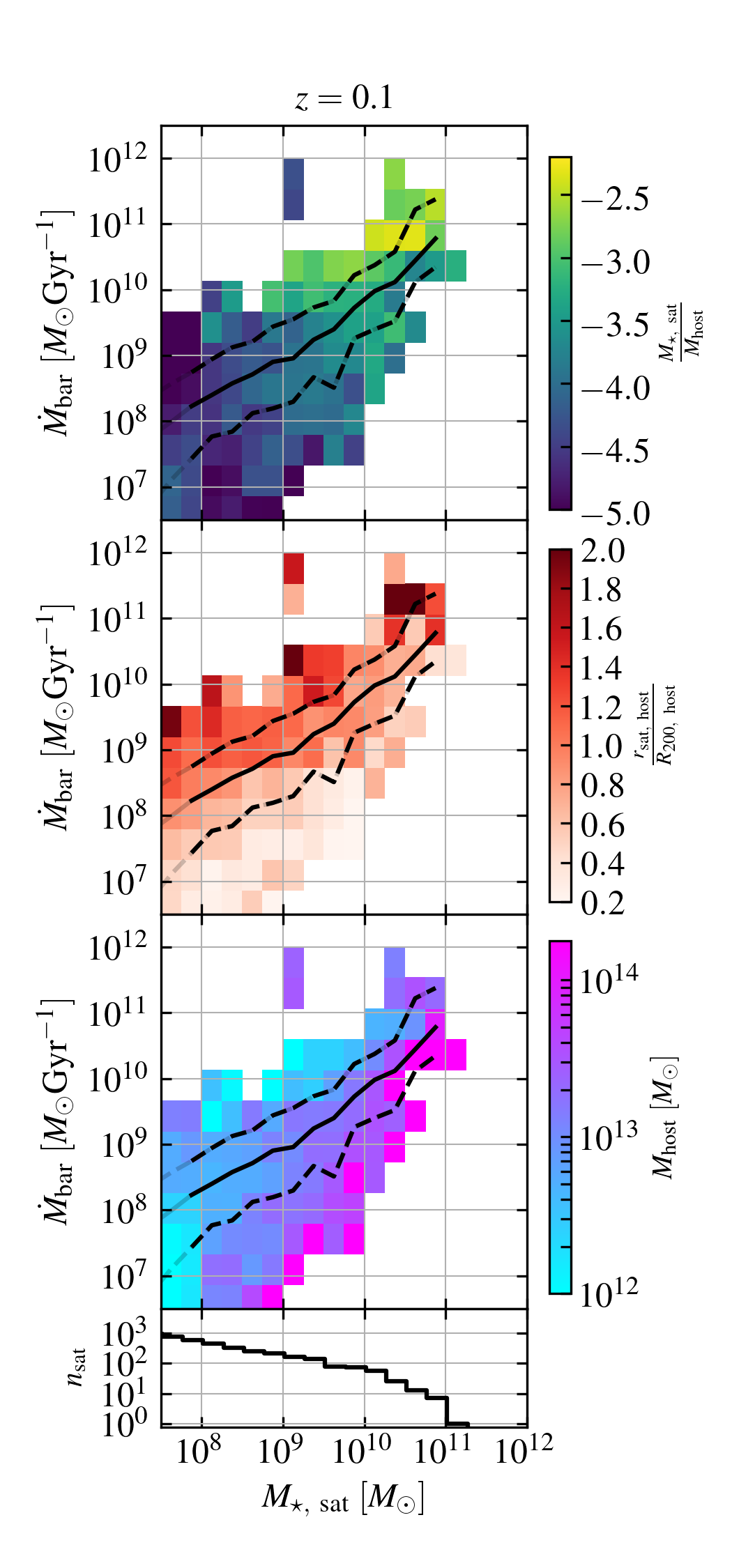}
\caption{Top panel: The median baryon accretion rate to satellite subhaloes as a function of their galaxy's stellar mass, coloured by the binned median stellar-halo mass ratio: $\frac{M_{\star,\rm\ sat}}{M_{\rm host}}$. Second panel: The median baryon accretion rate for satellite subhaloes as a function of their stellar mass, coloured by the binned median halo-centric distance of a satellite to its host's center-of-mass, normalised by the hosts virial radius: $(|{\bf r}_{\rm com,\ sat}-{\bf r}_{\rm com,\ host}|)/R_{\rm 200,\ host}$. Third panel: The median baryon accretion rate to satellite subhaloes as a function of their galaxy's stellar mass, coloured by the binned median host mass, $M_{\rm halo}$. Bottom panel: the number of satellite subhaloes included in each mass bin. Line transparency has been increased where the average efficiency has been calculated from a bin in which more than $50$\% of haloes were subject to an accretion flux of less than 50 gas particles. We use $20$ bins in stellar mass between $M_{\star}=10^{7}M_{\odot}$ and $M_{\rm star}=10^{12}M_{\odot}$. At fixed stellar mass, we see that the satellites with the greatest baryon accretion rates are, on average, (i) further from their host, (ii) of higher mass relative to their host,  and (iii) typically in less massive haloes.}
\label{fig:s3:subhalo}
\end{figure}

In this section, we briefly discuss accretion onto satellite subhaloes.  Many semi-analytic models of galaxy formation (e.g. \citealt{Lagos2018b,Lacey2016,Henriques2015}) assume that satellite subhaloes are instantaneously cut off from cosmological accretion as soon as they become satellite, subsequently impacting the star formation activity in satellite galaxies. This is an important assumption, and means that many satellite galaxies are very quickly and efficiently quenched in said models. With more realistic hydrodynamical simulations such as \eagle, it is possible to explore the validity of this assumption.

In the top three panels of Figure~\ref{fig:s3:subhalo}, we show the median baryon accretion rate for satellite subhaloes as a function of their stellar mass (the space coloured by different parameters). We note to the reader that the accretion rates to subhaloes that we quote are simply the summed mass of particles which entered a given subhalo between snapshots (irrespective of the origin of each particle). In the top panel, the parameter space is coloured by the binned median stellar-halo mass ratio $\left(M_{\star}/M_{\rm halo}\right)$. At fixed stellar mass, we see that on average, haloes with higher baryon accretion rates (than expected for their stellar mass) also have a higher stellar-halo mass ratio than their counterparts with lower baryon accretion rates. \citet{Wright2019} showed that the quenching timescales of satellite galaxies in \eagle\ were strongly correlated with the ratio between their stellar mass to their host halo mass. This tells us that the physical effect driving the lengthening of the quenching timescale in satellites is likely continuing gas accretion, which is larger in satellites that are massive relative to their host halo (leading to quenching timescales of $\gtrsim 5$~Gyr in the most massive satellites, see \citealt{Wright2019}).

In the second panel of Figure~\ref{fig:s3:subhalo}, the parameter space is coloured by the binned median halo-centric distance of a satellite to its host's center-of-mass, normalised by the host's virial radius: $(|{\bf r}_{\rm com,\ sat}-{\bf r}_{\rm com,\ host}|)/R_{\rm 200,\ host}$.  At fixed stellar mass, we see that the satellites with the greatest baryon accretion rates are, on average, further from their host than those subhaloes with relatively low baryon accretion rates. This tells us that baryon accretion onto the haloes of satellites appears to be suppressed when the satellites fall deep into the host's potential. 

In the third panel of Figure~\ref{fig:s3:subhalo}, the parameter space is coloured by the binned median host mass, $M_{\rm halo}$.  At fixed stellar mass, we see that the satellites with the greatest baryon accretion rates are more likely to exist in less massive haloes, while satellites in larger haloes ($\approx 10^{14}M_{\odot}$) typically experience depleted baryon accretion rates. Since we only measure gross accretion and do not account for outflows/gas loss, the reduced gas accretion rates does not directly imply the influence of environment, but rather that gas remains in the CGM or accretion occurs preferentially to the central galaxy in these circumstances.

While we find there are conditions under which satellite gas accretion can be suppressed, it is important to note that we find satellite subhaloes {\it can continue to accrete baryons}, and they are not completely cut-off from cosmological accretion (as many SAMs assume). This aligns well with the work of \citet{Hafen2020}, who show using the FIRE-2 simulations that a significant portion of CGM gas can be accreted to satellite galaxies. It is obvious that this has fundamental implications for the galaxies inhabiting these subhaloes, and their subsequent evolution. We defer a more complete discussion of accretion onto satellite subhaloes (specifically, the circumstances under which satellites can {\it net} accrete baryonic matter, and the detailed relationship between accretion and quenching) for a future paper.

\section{Accretion as a function of model physics}\label{sec:s4}
In this section, we build on \S \ref{sec:s3}, and focus on analysing the same topics when changes are made to feedback physics. In particular, we explore the qualitative and quantitative differences in accretion onto haloes that arise with the inclusion and parameterisation of stellar and AGN feedback. 

\subsection{Comparing baryon and dark matter accretion}
\label{sec:s4:baryondm}
\begin{figure*}
\includegraphics[width=1\textwidth]{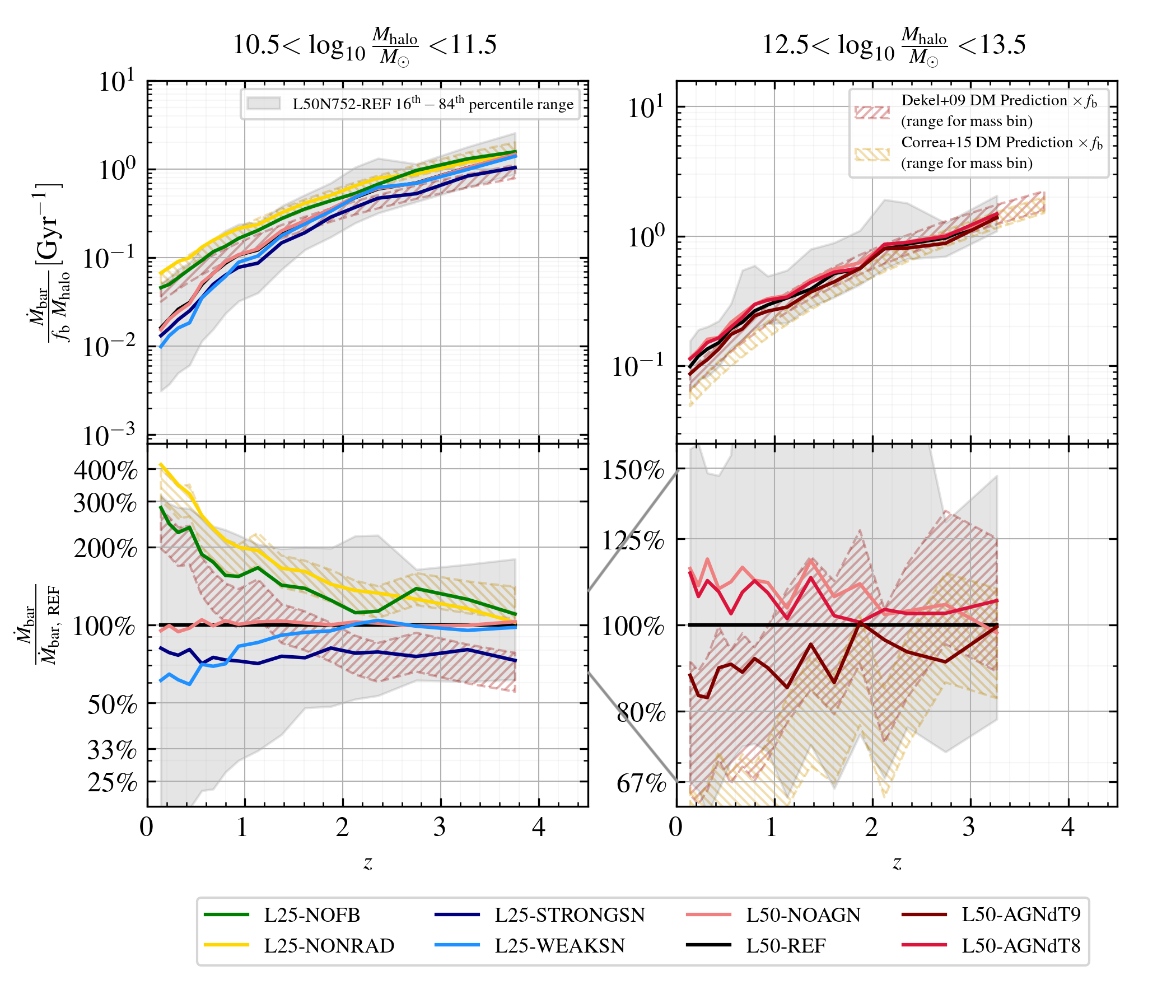}
\caption{Top panels: Median total (both smooth and merger-based) baryon accretion efficiencies in bins of halo mass (left: $10^{10.5}M_{\odot}<M_{\rm halo}<10^{11.0}M_{\odot}$, right: $10^{12.5}M_{\odot}<M_{\rm halo}<10^{13.5}M_{\odot}$) over the redshift range $z=0$ to $z=4$, for runs with various physical models, outlined in Tables~\ref{tab:sims} and \ref{tab:sims2} (\eagle\ reference physics in black, otherwise solid coloured lines). Bottom panels: The median total (both smooth and merger-based) baryon accretion efficiencies for each run, normalised by the baryon accretion efficiency of the L50-REF run. We display the $16^{\rm th}-84^{\rm th}$ percentile range of accretion efficiencies for the L50-REF run as the shaded grey region. Our work is also compared to the DM-based prediction presented in \protect{\citet{Dekel2009}} and \protect{\citet{Correa2015c}} (hatched pink and yellow, respectively, shaded regions for each mass range). We draw the reader's attention to the different axis ranges in each of the bottom panels. In the lower halo mass range (left) we focus on stellar feedback implementations. We see that compared to the L25-NONRAD and L25-NOFB runs, once stellar feedback is implemented (either in the L50-NOAGN run or the L50-REF run) there is a significant drop in baryon accretion onto haloes, particularly at low redshift (where we see differences of $\approx300-400\%$). In the higher halo mass range (right) we focus on AGN feedback implementations, where it is clear that removing AGN feedback or using a more efficient AGN feedback model can alter baryon accretion efficiencies at late times by $\approx 20-30\%$.} 
\label{fig:s4:literaturecomparison2}
\end{figure*}

\begin{figure*}
\includegraphics[width=1\textwidth]{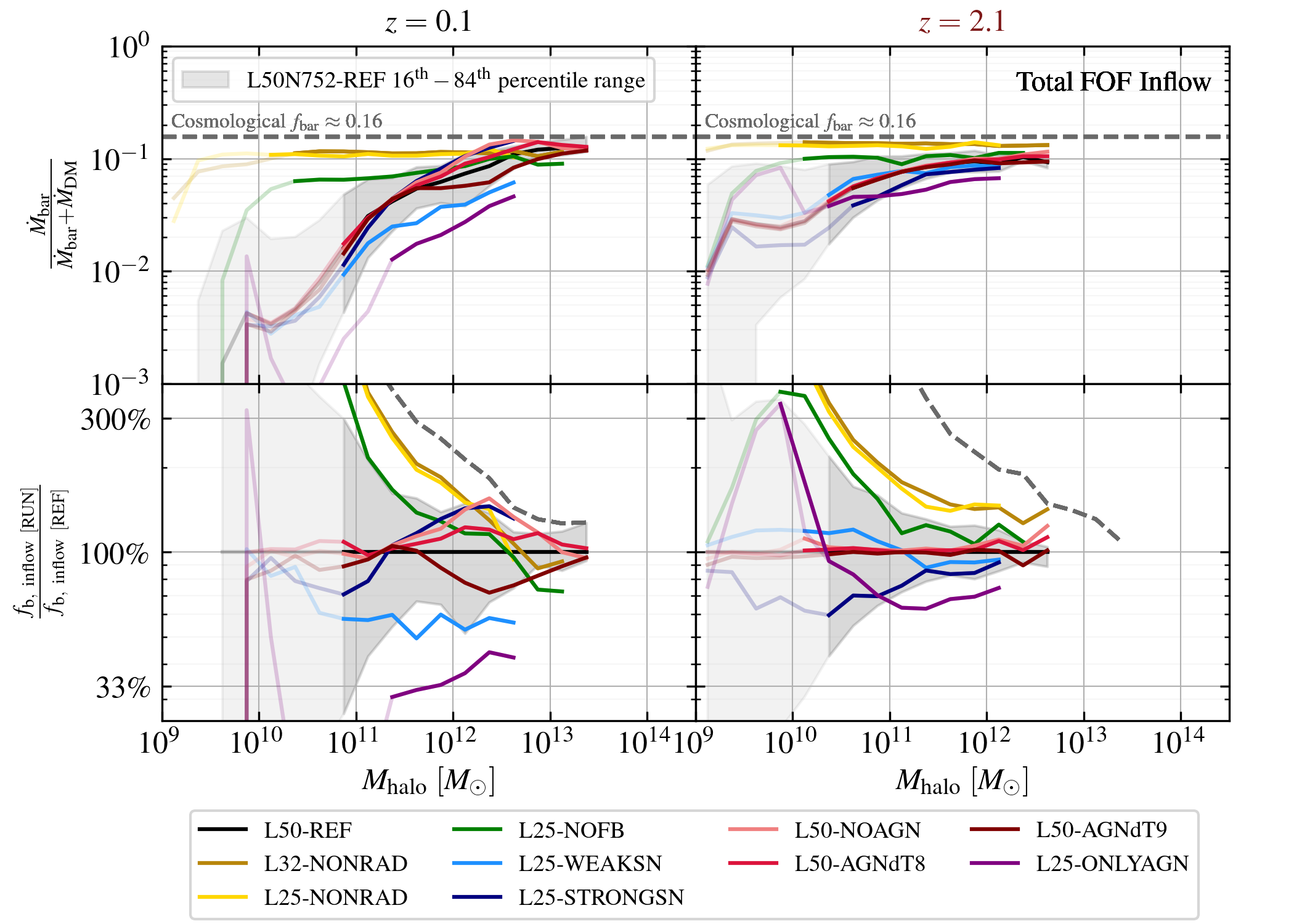}
\caption{Top panels: The baryon fraction of field halo-accreted matter as a function of halo mass for redshifts $z\approx0$ (left) and $z\approx2$ (right) for \eagle\ runs with various physical implementations, outlined in Table \ref{tab:sims} (reference physics in black, otherwise solid coloured lines). Bottom panels: The baryon fraction of field halo-accreted matter, normalised by the baryon fractions obtained in the L50-REF run. Line transparency has been increased where the average efficiency has been calculated from a bin in which more than $50$\% of haloes were subject to an accretion flux of less than 50 particles. To varying extents (and with different functional forms), we see a trend of increasing halo-accreted matter baryon fractions with halo mass for each \eagle\ run, with the exception of the adiabatic run which has a roughly constant inflow baryon fraction with halo mass. We see stellar feedback primarily  influences the low halo mass population, $M_{\rm halo}<10^{11.5}M_{\odot}$, while AGN feedback influences the high halo mass population, $M_{\rm halo}>10^{11.5}M_{\odot}$ (particularly at low redshift).}
\label{fig:s4:m200inflowfb}
\end{figure*}

In Figure~\ref{fig:s4:literaturecomparison2}, we show the median smooth baryon accretion efficiencies in $2$ bins of halo mass over the redshift range $z\approx0$ to $z\approx5$, for various feedback implementations as outlined in Tables~\ref{tab:sims} and \ref{tab:sims2}. This illustrate the main trends in gas accretion efficiency that we observe when changing feedback physics over cosmic time (which we explore further in Figure \ref{fig:s4:m200inflowfb}, comparing baryon and DM accretion rates directly). The top panels show the raw gas accretion efficiency over cosmic time, while the bottom panels illustrate the gas accretion efficiency at each redshift, normalised by the gas accretion efficiency seen in the L50-REF run. The left panels in Figure \ref{fig:s4:literaturecomparison2} focus on the halo mass range $10^{10.5}M_{\odot}<M_{\rm halo}<10^{11.5}M_{\odot}$ - the range in which we observe the largest differences with stellar feedback physics, compared to non-radiative and no-feedback physics. In these panels we illustrate the runs in which changes to stellar feedback (or lack thereof) have been made: L50-REF, L32-NONRAD (our in-house \gadget\ box), L25-NONRAD, L25-NOFB, L50-NOAGN, L25-WEAKSN and L25-STRONGSN). The right panels focus on the halo mass range $10^{12.5}M_{\odot}<M_{\rm halo}<10^{13.5}M_{\odot}$ - the range in which we observe the largest differences with varying AGN feedback physics. In these panels, we illustrate the runs in which changes to AGN feedback (or lack thereof) have been made: L50-REF, L50-NOAGN, L50-AGNdT8, and L50-AGNdT9. We also compare our work to the DM-based predictions presented in \citet{Dekel2009} (see Equation \ref{eq:2}), and \citet{Correa2015a,Correa2015b,Correa2015c}. These mass scales correspond to either side of the transition mass scale ($M_{\rm halo}\approx10^{12}M_{\odot}$) described in \citet{Bower2017} - above which star formation driven outflows cease to be buoyant, and are unable to prevent central mass build-up (which ultimately feeds BH mass accretion and AGN feedback). \\

Before looking at the influence of stellar and AGN feedback, we draw the reader's attention to the differences between the L25-NONRAD (yellow) run and the L25-NOFB (green) run in the left panels of Figure \ref{fig:s4:literaturecomparison2} - for the mass range $10^{10.5}M_{\odot}<M_{\rm halo}<10^{11.5}M_{\odot}$. We notice that while the non-radiative and no-feedback runs follow the same functional form over redshift (overpredicting baryon accretion rates compared to the full physics L50-REF run towards low redshift), the L25-NOFB run exhibits systematically lower baryon inflow rates than the L25-NONRAD run by a factor of $\approx 25\%$ (or $0.1$~dex) for $z\lesssim2$. We attribute this to the buildup of particles around the FOF boundary for haloes in the L25-NONRAD run, and the subsequent stochastic crossing and ejection from this boundary adding to accretion events. This is not the case in the L25-NOFB run, where we initially see higher accretion rates due to less thermal pressure with the inclusion of radiative cooling (at $z\gtrsim2.5$), but less buildup and stochastic FOF accretion events at later times. To verify this reasoning, we tested imposing our ``stability'' criterion: where we require particles to remain in the halo at the snapshot after accretion onto be considered a legitimate inflow candidate. When we impose this requirement, we observe that inflow rates in the L25-NOFB and L25-NONRAD runs were essentially identical, supporting our reasoning for the enhanced accretion in L25-NONRAD in Figure \ref{fig:s4:literaturecomparison2}. 

Moving towards the influence of stellar feedback, we find that there is tension between the \citet{Dekel2009} and \citet{Correa2015a,Correa2015b,Correa2015c} predictions and the L50-REF run at lower halo mass ($10^{10.5}M_{\odot}<M_{\rm halo}<10^{11.5}M_{\odot}$) particularly at late times, with our accretion efficiencies significantly ($\approx 0.6$~dex on average) lower than that those predicted (as also observed in Figure \ref{fig:s3:literaturecomparison1}. Since the accretion efficiencies in the L25-NONRAD, L32-NONRAD and L25-NOFB runs appear to agree very well with the DM-based predictions (within a factor of $\approx 2$), we argue that the implementation of baryonic feedback processes, in particular stellar feedback, is responsible for the tension at the low-mass end. Moreover, since we also observe the tension with DM-based predictions in the L50-NOAGN run, we argue that {\it it is specifically the inclusion of stellar feedback which causes baryon inflow onto  low mass haloes to be severely suppressed (by up to $\approx 1$~dex)}, agreeing with the arguments of \citealt{Voort2011a} with the older OWLS simulations.

The physical origin of this result is outlined in the work of \citet{Bower2017}, who show for haloes of mass $M_{\rm halo}\lesssim10^{12}M_{\odot}$ that star formation driven outflows are more buoyant than any tenuous corona surrounding the halo, ultimately suppressing gas inflow. This also agrees with the work of \citet{Mitchell2019}, who find that the halo-scale mass loading factor, $\eta=\langle \dot{M}_{\rm wind,\ halo} \rangle/\langle \dot{M}_{\star}\rangle$, is highest in this mass range, as a consequence of stellar feedback driven outflows. If we increase the energy injected by stellar feedback (as in the L25-STRONGSN run, dark blue), we see that gas accretion efficiencies are suppressed by $\approx 20-30\%$ compared to L50-REF for the full redshift range, tapering slightly towards $z=0$. \citet{Correa2018} illustrate a similar result at the {\it galaxy}-scale in the L25-STRONGSN run, where gas accretion was also suppressed for central galaxies with $M_{\rm gas}\lesssim10^{12}M_{\odot}$. Somewhat counter-intuitively, if we instead tune down the energy injection from stellar feedback (L25-WEAKSN, light blue), then we see a slight suppression in gas accretion rates towards $z=0$ (a result also found at the galaxy-scale by \citealt{Correa2018}). This highlights the importance of considering the interplay {\it between} stellar and AGN feedback - particularly how stellar feedback affects SMBH behaviour. When stellar feedback is weak, SMBHs in \eagle\ are allowed to grow more due to the accumulation of gas in galactic central regions, leading to stronger AGN feedback in low-mass haloes \citep{Bower2017}. As such, we argue that the slight decrease in gas accretion seen in L25-WEAKSN is driven by more efficient AGN relative to the L50-REF run.

Focusing now on the right hand panels (where we use a higher mass bin, $10^{12.5}M_{\odot}<M_{\rm halo}<10^{13.5}M_{\odot}$) we concentrate on AGN driven variations in accretion rates. The reader should note that we only test the influence of AGN feedback by using runs with altered temperature boost values, essentially controlling how ``explosive'' the AGN feedback is: meaning we do not consider alterations to the ``strength'' of AGN (i.e., the energy injection rate). Our work (with fiducial physics) and the DM-based predictions agree to within a factor of $2$ in this higher mass regime, and show the same functional form across redshift. This is contextualised by the results in \S \ref{sec:s3}, where we show with fiducial physics that baryon accretion rates approach DM accretion rates (scaled to a factor of $f_{\rm b}$) in higher mass haloes, $M_{\rm halo}\gtrsim10^{11.5}M_{\odot}$. Focusing on the bottom right-hand panel in Figure~\ref{fig:s4:literaturecomparison2}, we note that variations in AGN feedback make a small difference to the baryon accretion rate compared to the L50-REF box. At $z\approx0$ relative to the L50-REF run, in the L50-NOAGN run, we observe enhanced gas accretion efficiencies by $\approx 30\%$, and in the L50-AGNdT8 run (reduced AGN heating temperature from $10^{8.5}$~K in the reference box to $10^{8}$~K), we observe an enhancement of closer to $\approx\ 20\%$. Conversely, increasing the AGN heating temperature (to $10^{9}$K, L50-AGNdT9) induces a decrease in gas accretion efficiencies by $\approx 20\%$ at $z\approx0$. 

In both panels, we note that the influence of feedback in reducing gas inflow is most prevalent at late times. We believe this is the result of the higher binding energy of haloes at early times (a consequence of higher background matter density), which potentially causes feedback to be less effective at higher redshifts. This also provides context for the reduced influence of feedback at early times seen in Figure \ref{fig:s3:m200inflowfb}.

Thus, in \eagle, we see that the inclusion of stellar feedback can modulate accretion efficiencies in the halo mass range $10^{10.5}M_{\odot}<M_{\rm halo}<10^{11.5}M_{\odot}$ by a factor of $\approx 4$; while the inclusion and explosivity of AGN feedback can modulate accretion efficiencies in the halo mass range $10^{12.5}M_{\odot}<M_{\rm halo}<10^{13.5}M_{\odot}$ by a factor of $20\%-30\%$.  The much larger effect stellar feedback has relative to AGN feedback in these two regimes is due in part to the former acting on average on lower-mass haloes, whose potential wells are shallower than the massive haloes in which AGN feedback acts. We do not investigate the influence of AGN feedback on a halo-to-halo basis, however we refer the reader to the work of \citet{Davies2019} and \citet{Oppenheimer2020} who directly quantify the prevalence of SMBH activity in individual galaxies, and show that stronger SMBH activity is correlated with poor CGM gas content due to AGN-driven outflows. Another important consideration is the buoyant way in which stellar feedback-driven outflows expand inside haloes, which lead to significant halo gas mass being swept-up and removed from haloes. \citet{Mitchell2019} showed that halo-scale stellar feedback driven outflows in \eagle\ are characterised by a mass loading factor $\approx 10$ times greater than at the galaxy level.

In Figure~\ref{fig:s4:m200inflowfb}, similar to Figure~\ref{fig:s3:m200inflowfb}, we illustrate the baryon fraction of smooth-mode halo-accreted matter for redshifts $z\approx0$ and $z\approx2$ for our full collection of standard resolution \eagle\ variants, outlined in Table~\ref{tab:sims}, as well as our in-house \gadget\ L32-NONRAD run (Table~\ref{tab:sims2}). We also plot the ratio of $f_{\rm b,\ RUN}/f_{\rm b,\ L50-REF}$ in the bottom panels, to better illustrate the variations in each run compared to reference physics. We first focus on the effects of stellar feedback implementation, showing $f_{\rm b,\ inflow}$ values from the L50-REF model, as well as the weak (L25-WEAKSN, light blue) and strong (L25-STRONGSN, dark blue) stellar feedback runs. For halo masses $M_{\rm halo}\lesssim10^{11.5}M_{\odot}$, the L50-REF and L50-NOAGN runs (i.e., runs including stellar feedback) show a significant decrease in $f_{\rm b,\ inflow}$ compared to the L32-NONRAD, L25-NONRAD and L25-NOFB runs by up to $\gtrsim 300\%$, with the discrepancy increasing towards low redshift. 

At higher redshift ($z\approx2$, right panels), in the halo mass range $M_{\rm  halo}\lesssim10^{11.5}M_{\odot}$, we see that decreasing the strength of stellar feedback increases the baryon fraction of inflow by $\approx20\%-30\%$ compared to the L50-REF run, and that increasing the strength of stellar feedback decreases the baryon fraction of inflow by $\approx 40\%-50\%$ compared to the L50-REF. Interestingly, above this mass range, we see a very slight ($10\%-20\%$) decrease in  $f_{\rm b,\ inflow}$ for both the L25-STRONGSN and L25-WEAKSN boxes. Here, we again begin to note the importance of understanding the interplay between stellar and AGN feedback in order to accurately interpret these results (see \citealt{Bower2017} for an in-depth analysis of how stellar feedback modulates AGN activity in \eagle). In the case of the L25-STRONGSN box, we attribute the drop in baryon accretion directly to the stronger stellar feedback (similar to the lower halo mass range).  In the L25-WEAKSN box (where one might expect baryon inflow onto be bolstered, rather than suppressed), we attribute the drop to enhanced AGN feedback, in the absence of sufficiently strong stellar feedback. Physically, less pressure support and heating from stellar feedback allows more gas to be funnelled towards the center of haloes which are host to a SMBH. This increases the BH accretion rate (and, therefore, the energy injection rate from AGN feedback),  causing strong outflows in haloes that in the reference run were only affected by stellar feedback. The extreme case is shown in the L25-ONLYAGN run, where the lack of stellar feedback appears to increase the ability of AGN feedback to suppress gas inflow at low halo masses (down to $M_{\rm halo}\approx10^{10.5}M_{\odot}$).

Moving towards lower redshift ($z\approx0$), we see that the differences between the L25-STRONGSN and L50-REF runs in the mass range $M_{\rm halo}\lesssim10^{11.5}M_{\odot}$ are similar to those analysed above at $z\approx 2$. This said, at $z\approx0$, we actually measure enhanced baryon accretion in the L25-STRONGSN box compared to L50-REF for haloes of mass $M_{\rm halo}\gtrsim10^{11.5}M_{\odot}$. Interestingly, in this mass range, the behaviour of the L25-STRONGSN run and the L50-NOAGN run appear almost identical. We argue that this is due to the increased energy injection from stellar feedback causing SMBHs to have decreased accretion rates, eventually leading to suppressed AGN feedback. Hence, in this mass range, doubling  stellar feedback energy injection induces the same baryon accretion behaviour we would expect if there were no AGN feedback whatsoever. \citet{Correa2018} show that in the L25-STRONGSN run the gas accretion at the galaxy-scale increased relative to the L50-REF run for galaxies with $M_{\rm gas}\gtrsim10^{12}M_{\odot}$. This again points to variations in the gas accretion rate at the galaxy-scale to be largely driven by the variations in the baryon content of the matter accretion at the halo level.

Focusing instead on the effects of varying the AGN feedback implementation (comparing L50-REF, L50-NOAGN, L50-AGNdT8 and L50-AGNdT9), we see for both redshift panels that there is little change below $M_{\rm halo}\approx10^{11.5}M_{\odot}$.  At $z\approx2$, we see differences between these runs only for the highest mass haloes, $M_{\rm halo}\gtrsim10^{12}M_{\odot}$. Omitting AGN feedback (L50-NOAGN, pink) may modestly increase baryon accretion by $\approx10\%-20\%$ compared to the L50-REF run, and decreasing the injection temperature to $10^{8}$K (from $10^{8.5}$K in the reference physics model) in the L50-AGNdT8 run (red) has the same effect of modestly increasing the baryon inflow (albeit to a slightly lesser degree). Increasing the injection temperature to $10^{9}$K in the L50-AGNdT9 run (maroon) decreases the baryon inflow efficiency by the same very modest percentages.

Towards $z\approx0$, the same variations in AGN treatment produce much larger variations in baryon inflow. We see the largest amplitude of modulation in the halo mass range $10^{12}M_{\odot}\lesssim M_{\rm halo}\lesssim10^{13}M_{\odot}$ - where the omission of AGN feedback increases $f_{\rm b,\ inflow}$ relative to L50-REF by $\approx70\%-80\%$; decreasing the AGN injection temperature increases $f_{\rm b,\ inflow}$ by $\approx30\%-40\%$ compared to L50-REF; and increasing the AGN injection temperature decreases $f_{\rm b,\ inflow}$ by $\approx 30\%$. Interestingly, for the highest mass haloes in the $50$Mpc boxes at $M_{\rm halo}\gtrsim10^{13}M_{\odot}$, AGN physics seems to have very little impact on baryon accretion. \citet{Lagos2018a} show that high mass centrals ($M_{\star}\gtrsim10^{11.8}M_{\odot}$, corresponding to haloes of mass of $M_{\rm halo}\gtrsim10^{13}M_{\odot}$) seem to be preferentially more star-forming and fast-rotating than observational counterparts, indicating that AGN feedback in the most massive \eagle\ systems may not be strong enough. Our findings provide context for this result, with there being little modulation of halo-scale inflow by AGN feedback in the L50-REF run compared to the L50-NOAGN run for massive haloes, $M_{\rm halo}\gtrsim10^{13}M_{\odot}$. 

 Our results show that global AGN-driven suppression of {\it halo-scale} gas inflow is relatively mild, of order $\approx 30\%$. If we compare to the {\it galaxy-scale} accretion rates presented in \citealt{Correa2018} (which show a strong AGN modulation of inflow rates by $\approx 0.5$~dex in the halo mass range $10^{12}M_{\odot}\lesssim M_{\rm halo}\lesssim10^{12.5}M_{\odot}$), the picture that emerges is one where AGN activity greatly influences gas behaviour at the CGM-scale to ultimately suppress gas inflow to galaxies

\subsection{Channels of baryon and DM inflow}
\label{sec:s4:channels}
\begin{figure*}
\includegraphics[width=0.86\textwidth]{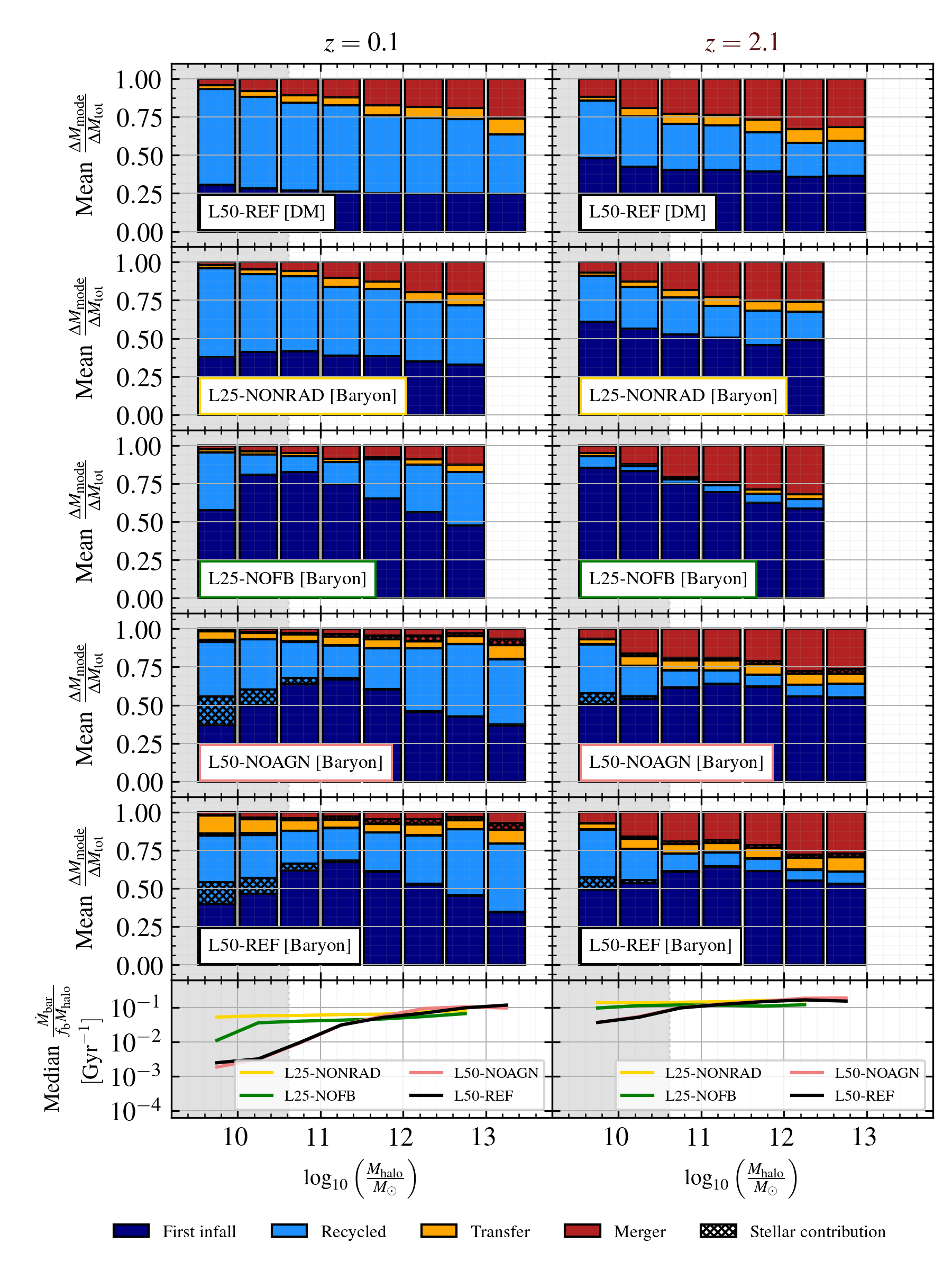}
\caption{Channels of baryon and DM accretion as a function of halo mass for $z\approx0$ (left panels) and $z\approx2$ (right panels). We use the same definitions of accretion channels as outlined in \S \ref{sec:methods:chumm} and Figure \ref{fig:s3:accretionbreakdown}, and the same halo mass bins as in Figure \ref{fig:s3:accretionbreakdown}. In the first row of panels we present our DM results from L50-REF for comparison (with DM channels not significantly changing between runs). In the second-fifth rows of panels, we present the baryon accretion channels for $4$ \eagle\ variants, from top to bottom: (i) L25-NONRAD, (ii) L25-NOFB, (iii) L50-NOAGN, and (iv) L50-REF. The bottom row of panels shows the median gas accretion efficiency in each of these bins, to remind the reader of the relative normalisation of raw accretion rates as a function of halo mass for these different runs. We show that the sequential introduction of radiative cooling, stellar and AGN feedback influences inflow channels considerably relative to DM-only physics.}
\label{fig:s4:accretionbreakdown}
\end{figure*}


We now move on to discuss the effect that baryonic physics has on the breakdown of accretion channels across cosmic time. Figure~\ref{fig:s4:accretionbreakdown} shows the breakdown of accretion channels for baryons and DM in a selection of the \eagle\ variations as a function of halo mass, at $z\approx0$ and $z\approx2$. In the first row of panels, we show the DM accretion channels from L50-REF, noting that the breakdown of DM accretion channels were effectively identical for each run. From the second row of panels down, show baryon accretion channels for (i) L25-NONRAD, (ii) L25-NOFB, (iii) L50-NOAGN, and (iv) L50-REF - effectively step-wise introducing new baryonic physics. The bottom panels act to remind the reader of the normalisation of baryon accretion efficiencies in each mass bin for the various runs. 

Starting with the non-radiative physics run (L25-NONRAD, 2nd row of panels), we see that the breakdown of baryon accretion is very similar to that of DM in L50-REF for each of the halo mass bins. The rate of mergers is highest for massive haloes at high redshift, approaching $35-40$\% of all mass accretion at $z\approx2$. The proportion of recycled accretion increases at late times to a steady $\approx40-50\%$, decreasing modestly with halo mass where mergers play a more dominant role. Transfer of baryons appears significant even in the absence of feedback in L25-NONRAD, which we hypothesise is the result of (i) unstable accretion of tenuous particles and subsequent ejection by the pressure-supported atmosphere of haloes in this run, and/or (ii) cosmic-web stripping \citep{Benitez2013,Beckmann2017}. In the 3rd row of panels, we show the influence of introducing radiative cooling on the breakdown of accretion channels with the L25-NOFB run. This change alters the picture significantly, most notably reducing recycling in haloes with mass $M_{\rm halo}\gtrsim10^{11}M_{\odot}$ (particularly at $z\approx2$). It is this change that appears to produce the peak in first infall baryon accretion that we previously saw with reference physics in Figure \ref{fig:s3:accretionbreakdown}, where recycling would otherwise dominate. This drastic change tells us that the enhanced recycling seen in L25-NONRAD is a result of non-radiative physics. 

While we have not directly investigated particle trajectories, in the non-radiative runs we believe that more thermal pressure around the FOF boundary means that particles may enter the halo and promptly ``bounce'' back outside the boundary. When these particles are eventually re-accreted, this second accretion will be considered recycled. With radiative cooling, the pressure support at the halo boundary is reduced, meaning that first infall particles are more likely to stay in the halo, and not be re-accreted at a later time. At at $z\approx 2$ in L25-NOFB, baryons are slightly more likely to be accreted via mergers (i.e., less likely to be smoothly accreted) compared to DM in higher mass halos ($M_{\rm halo}\gtrsim10^{11.5}M_{\odot}$). This is naturally expected in a run implementing radiative cooling, as these processes allow baryons to accumulate at the centres of haloes in a run-away fashion - also known as overcooling \citep{White1978} - increasing the amount of baryon mass contributed by mergers.

In the 4th row of panels, we introduce stellar feedback in a larger box with the L50-NOAGN run. An interesting result here is that the baryon growth from mergers drops by a factor of $\approx2$ across halo mass compared to L25-NOFB at $z\approx0$, producing the final small merger contribution we saw in Figure \ref{fig:s3:accretionbreakdown}. As discussed in \S \ref{sec:s3:baryondm:channels} and \ref{sec:s4:baryondm}, this is a consequence of stellar feedback suppressing gas inflow onto  low mass haloes (see bottom panels), causing them to be baryon-poor relative to the universal $f_{\rm b}$. Subsequent mergers involving these baryon-poor haloes provide predominantly DM mass, and minimal baryonic matter. The recycled component in L50-NOAGN remains suppressed compared to DM, but is slightly increased compared to baryons in L25-NOFB - a direct result of the ejection and re-accretion of baryons driven by stellar feedback. We observe a small increase in the transfer channel of accretion in L50-NOAGN compared to L25-NOFB, which we attribute to particles being ejected by particularly strong stellar feedback events and later re-accreted to un-related haloes. It appears that high-mass halos, $M_{\rm halo}\gtrsim10^{12}M_{\odot}$, are slightly more likely to see enhanced transferred accretion, a likely result of their deeper potential well and ability to attract particles previously ejected from smaller haloes, where stellar feedback-driven mass loading is largest \citep{Mitchell2019}. 

Finally, in the 5th row of panels, we introduce AGN feedback with L50-REF to reach the results we presented in \S \ref{sec:s3:baryondm:channels}. There are minimal changes to observe with reference physics comparing to L50-NOAGN, however, we see a slight change in that baryons are slightly less likely to be recycled and more likely to be transferred relative to L50-AGN at $10^{12}M_{\odot}\lesssim M_{\rm halo}\lesssim10^{13}M_{\odot}$ for the $z\approx0$ panel. This is due to AGN feedback boosting the outflow rates in massive haloes by factors of $3-10$ (with the exact value increasing with decreasing redshift; \citealt{Mitchell2019}). Because the momentum and energy injection onto outflows by AGN feedback is $\gtrsim 3$ times larger than stellar feedback at $z<1$ in these massive haloes for the \eagle\ reference physics \citep{Mitchell2019}, the ejected gas particles can move further away from the potential well of these massive haloes. A fraction of those particles are likely to escape the potential well and not make its way back to re-accretion. Without AGN feedback, this does not happen, and effectively all ejected particles from haloes are expected to be re-accreted by the same halo. Additionally, the higher virial temperature associated with massive haloes prevents reincorporation of AGN-ejected gas due to (i) direct thermal support in the intra-cluster medium (ICM), and (ii) less efficient cooling (given halo virial temperatures are above the peak of the radiative cooling curve).  We remark that the inter-halo transfer channel is further increased in the L50-AGNdT9 run (not shown here), where more explosive energy injection likely allows a larger proportion of particles to escape the potential of their host halo, and subsequently re-accrete to unrelated haloes. 

\section{Implications for modelling galaxy evolution}\label{sec:discussion}
In this section, we explore the implications of our results in \S~\ref{sec:s3} and \S~\ref{sec:s4} on our understanding of galaxy evolution, and their application to SAMs. We first discuss the dependence of our results on the model used. While we have established that there is significant feedback-induced suppression of halo-scale gas inflow onto low-mass haloes in \eagle, the scale at which inflows interact with outflows and the dynamics of this interface are not well understood. 

\citet{Mitchell2019} compare outflow rates and mass loading on various scales in \eagle\ with other hydrodynamical simulations (e.g. from Illustris-TNG - \citealt{Nelson2019}; Horizon-AGN - \citealt{Beckmann2017}). They find that although trends across halo mass are qualitatively similar, the biggest uncertainty comes from the scale over which outflows are measured. At $M_{\star}\approx10^{9}M_{\odot}$, outflow rates in Illustris-TNG appear to be stifled between $10$ and $50$ kpc, while outflow flux remains the same in \eagle\ at $50$ kpc compared to $10$ kpc. In general, Illustris-TNG seems to prefer a galactic-fountain picture with strong recycling over smaller scales. \eagle, on the other hand, produces outflows that remove comparatively less material from galaxies, but that become significantly entrained as they expand in the haloes (resulting in the ejection of material to much larger radii compared to Illustris-TNG, \citealt{Mitchell2019}). As such, it is important for us to note that simulation-based halo-scale inflow measurements, and the impact of sub-grid physics, are almost certainly strongly model-dependent (as is explored further in \citealt{Mitchell2020}). Similarly, variations in the baryon inflow rate onto haloes are also model-dependent. This highlights the need for observational constraints on the physical properties of accreting gas (temperature, density and metallicity), and in an upcoming paper (Wright et al. in prep), we show how these physical properties are sensitive to both the sub-grid physics implemented, and the channel through which the gas is accreting by. These differences may be a promising way of using observations to indirectly constrain the effect of outflows and inflows at the scale of haloes and beyond.
    
Moving onto the implications of our results for SAMs, we first remind the reader that most SAMs use merger trees produced from DM-only simulations as a base to form and evolve galaxies, with haloes growing their baryonic mass based on the net change in DM mass (modulo a factor of $f_{\rm b}$). The baryon budget within the halo is then managed analytically, and evolved with the specific models implemented within the SAM. Our work highlights a seldom-addressed issue comparing mass growth between SAMs and hydrodynamical simulations, with the significance of DM recycling indicating that gross DM accretion rates do not necessarily reflect net DM growth within a halo. One could argue that it is more appropriate to model baryon growth based only on first infall DM accretion rates (which would require halo-by-halo accretion calculations compared to a simple $\Delta M$ calculation). We defer a full discussion of this topic to a future paper.

While halo-scale inflow has been shown to be less sensitive to baryonic feedback compared to galaxy-scale inflow (e.g. \citealt{Voort2011a}, \citealt{Nelson2015}), we argue that the suppression of halo gas inflow with the inclusion of stellar feedback (by up to $\approx 1$~dex for lower mass haloes, $<10^{11}\,\rm M_{\odot}$, see Figure~\ref{fig:s4:m200inflowfb}) is significant enough to have implications on the treatment of baryons in SAMs. Most SAMs implement some form of stellar and AGN feedback by ejecting gas from galaxies into the CGM, or from the halo entirely. We have shown here, however, that feedback plays a dual role in not only ejecting baryons from galaxies, but also {\it preventing subsequent accretion onto their host halo}. We would expect this effect to be significant in haloes in the mass range  $M_{\rm halo}\lesssim10^{12}M_{\odot}$ (corresponding to $M_{\star,\ \rm cen}\lesssim10^{10}M_{\odot}$), where baryon inflow is $\lesssim50\%$ of that expected from DM (at $z\lesssim1$). Our resolution with L50-REF limits us to only making this conclusion for haloes of mass $M_{\rm halo}\gtrsim10^{10.5}M_{\odot}$ (corresponding to $M_{\star,\ \rm cen}\approx10^{7.5}M_{\odot}$), however we can confidently see this suppression continuing below this mass (to $M_{\rm halo}\approx10^{10}M_{\odot}$, where $M_{\star,\ \rm cen}\approx10^{7}M_{\odot}$) in the L25N752-RECAL run; see Appendix Figure \ref{fig:appendix:res}. 

In the aforementioned mass range ($M_{\rm halo}\lesssim10^{12}M_{\odot}$) that we have established there to be baryon depleted inflow, haloes in current SAMs are subject to over-predicted gas inflow rates by up to 1~dex. \citet{Mitchell2018} showed that halo baryon fractions in \eagle\ are far lower than those found in {\sc GALFORM} for haloes with mass $\lesssim10^{12.5}M_{\odot}$. Our findings contextualise this result, showing that these overly baryon rich haloes in SAMs could be plausibly explained by over-predicted baryon growth. This puts pressure on feedback mechanisms in SAMs, particularly stellar feedback, to remove the excess baryons in order to reproduce observational standards. While SAMs have had success in reproducing a number of calibrations from observations (e.g. the stellar mass function, see \citealt{Henriques2015,Lacey2016,Lagos2018b}), it has been shown that they do so for very different reasons compared to hydrodynamical simulations \citep{Mitchell2018}. 

\citet{Lu2017} show, with a semi-analytic approach (outlined in \citealt{Lu2016}), that an exclusively ``ejective''-mode stellar feedback implementation cannot simultaneously recover the mass-metallicity relation (MMR) and stellar mass function found in low mass Milky Way dwarfs. They argue that with the strong stellar-driven outflows required to suppress star formation also comes the ejection of metals, eventually producing an artificially steep MMR that predicts lower metallicity dwarfs than observed. On the other hand, hydrodynamical simulations, which naturally include the ``preventative'' effect of feedback and suppression of gas inflow (not just the ``ejective'' effect), are able to reproduce the shape of the observed MMR fairly robustly where inner-haloes retain adequate metal content (e.g. \eagle\ - \citealt{Schaye2015}; Illustris - \citealt{Torrey2014}). \citet{Agertz2020} highlight the ability of the MMR in assessing the accuracy of models, showing that changes to stellar feedback significantly influence the MMR without necessarily influencing other commonly used scaling relations (e.g. involving half-mass radii and stellar velocity dispersions). Our results further support the notion that ``preventative''-mode feedback (in particular, stellar feedback-induced suppression of baryon inflow), in addition to traditional ``ejective''-mode feedback, is necessary to accurately model galaxy formation and evolution in SAMs: not just at the galaxy-scale, but also at the halo-scale. 

In the past, SAMs have had trouble reproducing the expected X-ray scaling relationships of $L^{*}$ galaxies \citep{Crain2010}, and groups/clusters \citep{Bower2006,Bower2008}. \citealt{Bower2008} present a version of the {\sc galform} SAM \citep{Bower2006} including ``radio-mode'' AGN-induced ICM heating, acting in an ejective and subsequent preventative manner to reduce the baryon fraction of haloes. This feedback implementation allows X-ray scaling relations to be better reproduced, illustrating the importance of reducing the baryon content of halos.

Lastly, as briefly discussed in \S \ref{sec:s3:subhaloes}, we show that satellite subhaloes are certainly still capable of accreting cosmological gas. Most SAMs will assume that satellite galaxies are cut off from cosmological accretion upon infall to a larger host halo: an assumption which we plan to further investigate in a future paper, with an analysis of the conditions under which satellite subhaloes are able to {\it net} accrete baryonic matter.

The above discussion points to the fact that feedback has two-fold profound influence on halo-scale baryonic mass assembly - in (i) removing baryonic matter from galaxies and haloes, and then (ii) preventing further baryonic inflow. This level of complexity (particularly the preventative aspect) is not captured by SAMs, and important revisions are therefore required for these tools to accurately model the effects that we analysed in this paper.

\section{Summary}\label{sec:conclusions}
In this paper we have presented measurements of halo-scale baryon and DM inflow rates in the \eagle\ suite of simulations. Our FOF-based algorithm, \chumm, recovers gas inflow efficiencies ($\dot{M}_{\rm gas}/f_{\rm b}M_{\rm halo}$) that are in broad quantitative agreement presented in previous literature using hydrodynamical galaxy formation simulations (e.g. \citealt{Voort2011a}, \citealt{Correa2018}) to within $\approx 0.2$~dex at both $z\approx0$ and $z\approx2$ (see Figure~\ref{fig:s3:literaturecomparison1}). We consider accretion in 4 different channels: (i) first infall (accreting particles which have never been identified as part of a halo in the past), (ii) recycled-mode (particles which have been processed in a progenitor halo beforehand, but were most recently accreted from the field) (iii) transfer (particles which have been processed in a non-progenitor/unrelated halo beforehand, but were most recently accreted from the field) , and (iv) merger-mode (particles which were accreted to a halo which, at the previous snapshot, were part of another halo).
Our main findings using the fiducial physics \eagle\ run L50-REF were presented in \S~\ref{sec:s3}, and are summarised below:

\begin{itemize}
    \item Baryon accretion is suppressed relative to DM in low mass field haloes ($M_{\rm halo}\lesssim10^{12}M_{\odot}$, corresponding to $M_{\star,\ \rm cen}\lesssim10^{10}M_{\odot}$) by up to 1~dex (see Figure~\ref{fig:s3:m200inflowfb}). This effect is particularly prominent towards late times, where the halo-to-halo variance in $f_{\rm b,\ inflow}=\dot{M}_{\rm bar}/(\dot{M}_{\rm bar}+\dot{M}_{\rm DM})$ is also largest. In the same mass range, the baryon fractions of halo-scale inflow are very well-correlated with halo-wide baryon fractions (most notably in the CGM), suggesting a causal effect (see Figures \ref{fig:s3:inflowfbhalofb} and \ref{fig:s3:m200halofb}).
    \\
    \item Regarding accretion channels: 
    \begin{enumerate}
        \item The merger-mode growth channel is of lesser importance for baryons compared to DM in more massive haloes ($M_{\rm halo}\gtrsim10^{12}M_{\odot}$), particularly at late times. This is directly attributable to the baryon depletion of the low mass haloes that contribute mass in mergers (see Figure \ref{fig:s3:accretionbreakdown}). Baryon depletion of low mass haloes could either be due to the aforementioned baryon poor inflow, or stripping \& pre-processing in the case of satellites, see \citealt{Bahe2015,Bahe2019}), which are only capable of contributing significant mass in the form of DM during merger events.
        \item  Compared to baryonic matter, smoothly accreted DM is significantly more likely (by a factor of $\approx 2$) to have been recycled in a halo than to be on first infall  (with the exception of higher mass haloes, $M_{\rm halo}\gtrsim10^{12.5}M_{\odot}$, at $z\approx0$). This is a reflection of the ubiquitous suppression of baryon inflow onto low mass haloes over cosmic time: any given baryon particle is less likely to have been accreted at some point in the simulation compared to a DM particle.
    \end{enumerate}
    \item Inflow rates to satellite subhaloes are influenced by their stellar-to-halo mass ratio ($M_{\star}/M_{\rm halo}$) and their normalised halo-centric distance ($R_{\rm sat,\ halo}/R_{200,\rm\  halo}$) with more massive satellites and satellites at larger radii relative to their host being more likely to see higher gas accretion rates (Figure \ref{fig:s3:subhalo}). 
\end{itemize}

In \S \ref{sec:s4}, we go on to investigate the influence of sub-grid physics on the results that we summarise above, by making use of several additional simulation runs (outlined in Tables \ref{tab:sims} and \ref{tab:sims2}). We find that: 
\begin{itemize}
    \item In \eagle, the significant suppression (by up to $\approx1$ dex) of  $f_{\rm b,\ inflow}$ in low mass haloes ($M_{\rm halo}\lesssim10^{12}M_{\odot}$) compared to the universal $f_{\rm b}$ is a direct result of the introduction of stellar feedback (see Figures \ref{fig:s4:literaturecomparison2}, \ref{fig:s4:m200inflowfb}). In addition, the interplay between stellar and AGN feedback in \eagle\ has a profound influence on gas inflow. Weakening stellar feedback can lead to runaway SMBH accretion and subsequent AGN outflow, which can heavily suppress gas accretion rates (for an in-depth explanation of the interaction between stellar and AGN feedback in \eagle, see \citealt{Bower2017}). For a fixed stellar feedback implementation, we see that altering AGN feedback injection temperature (``explosivity'') can modulate gas inflow rates in high mass haloes ($M_{\rm halo}\gtrsim10^{12.5}M_{\odot}$ by $\approx25-50\%$) - see Figures \ref{fig:s4:literaturecomparison2} and \ref{fig:s4:m200inflowfb}.
    
    \item Gas inflow channels are strongly influenced by sub-grid physics (Figure \ref{fig:s4:accretionbreakdown}). Gas and DM are accreted onto haloes with similar contributions from both smooth/merger and recycled/first infall channels in the runs with no radiative cooling, star formation and feedback. Introducing radiative cooling leads to a decrease in the gas recycling fractions, which we attribute to gas cooling removing the thermal pressure at the virial radius that naturally builds up in the non-radiative runs. Adding stellar feedback with the L50-NOAGN run increases the importance of the baryon recycling and transfer accretion channels at all halo masses as a result of supernovae-driven outflows, which are eventually re-accreted onto the halo - see Figure \ref{fig:s4:accretionbreakdown}. Introducing AGN feedback with L50-REF has little influence on the importance of the different accretion channels analysed here compared to the L50-NOAGN run, however marginally reduces the recycling-mode and increases inter-halo transfer-mode of baryon inflow for high mass haloes, $M_{\rm halo}\gtrsim10^{12}M_{\odot}$, which we attribute to AGN outflows being sufficiently strong to entirely eject gas particles from haloes, without eventual reincorporation.
\end{itemize}

In \S \ref{sec:discussion}, we discuss the model-dependence of our results, and the implications that our findings have on semi-analytic approaches with supporting literature. We argue that preventative-mode feedback (i.e., suppressing gas inflow), in addition to traditional ejective-mode feedback (i.e., energetic gas outflows) are both important manifestations of stellar feedback in \eagle. This dual effect (even at the halo-scale) seems to be required to accurately model the behaviour of low mass haloes (with a particularly useful calibration of stellar feedback being the mass-metallicity relation - see \citealt{Lu2017,Agertz2020}). 

We point the reader towards the work of \citet{Mitchell2020} for an in-depth analysis of the physics of accretion in \eagle\ towards the CGM- and galaxy-scale, together with the interplay between outflows, inflows and recycling. In a future paper, we will look to further investigate gas accretion onto subhaloes in hydrodynamical simulations: specifically, the conditions under which subhaloes are able to {\it net} grow in baryonic mass. This is a topic of of particular interest for SAMs, as most will assume that satellite galaxies are abruptly cut off from cosmological accretion upon infall to a host. Additionally, in another future paper, we plan to examine the properties of accreting baryons at the halo-scale: specifically the temperature, density, and metallicity of inflow particles around the virial radius. We will investigate the sensitivity of these properties to sub-grid physics implementations, which could have observational applications in being able to constrain the effect of outflows and inflows at the halo-scale. 

\section*{Acknowledgements}

%
We thank the anonymous referee for their insightful report that helped improve the clarity of this paper. We also thank Jon Davies and Rob Crain for fruitful discussions.
RW is funded by a Postgraduate Research Scholarship from the University of Western Australia (UWA). CL is funded by the ARC Centre of Excellence for All Sky Astrophysics in 3 Dimensions (ASTRO 3D), through project number CE170100013. CL also thanks the MERAC Foundation for a Postdoctoral Research Award. CP acknowledges the support ASTRO 3D.
PM acknowledges a UWA Research Collaboration Award 2018 for funding his visit to UWA. 
This work made use of the supercomputer OzSTAR, which is managed through the Centre for Astrophysics \& Supercomputing, and maintained by Swinburne ITS. Their supercomputing program receives continued financial support for operations from Astronomy Australia Limited and the Australian Commonwealth Government through the National Collaborative Research Infrastructure Strategy (NCRIS). The \eagle\ simulations were performed using the DiRAC-2 facility at Durham, managed by the ICC, and the PRACE facility Curie based in France at TGCC, CEA, Bruyeres-le-Chatel.

\section*{Data Availability}

Particle data from a subset of the \eagle\ runs used for our analysis is publicly available at \url{http://dataweb.cosma.dur.ac.uk:8080/eagle-snapshots/} - specifically the L25-REF, L50-REF, L50-NOAGN, L50-AGNdT8, L50-AGNdT9, L25-WEAKSN, L25-STRONGSN, and L25N752-RECAL runs. All \velociraptor-generated halo catalogues, and  \treefrog\ merger trees, are available upon request from the corresponding author (RW). The code we used to generate accretion rates to haloes is available at \url{https://github.com/RJWright25/CHUMM}.




\bibliographystyle{mnras}
\bibliography{main.bib} 


\appendix

\section{Field halo mass distributions}\label{sec:appendix:massfunctions}
\begin{figure*}
\includegraphics[width=1\textwidth]{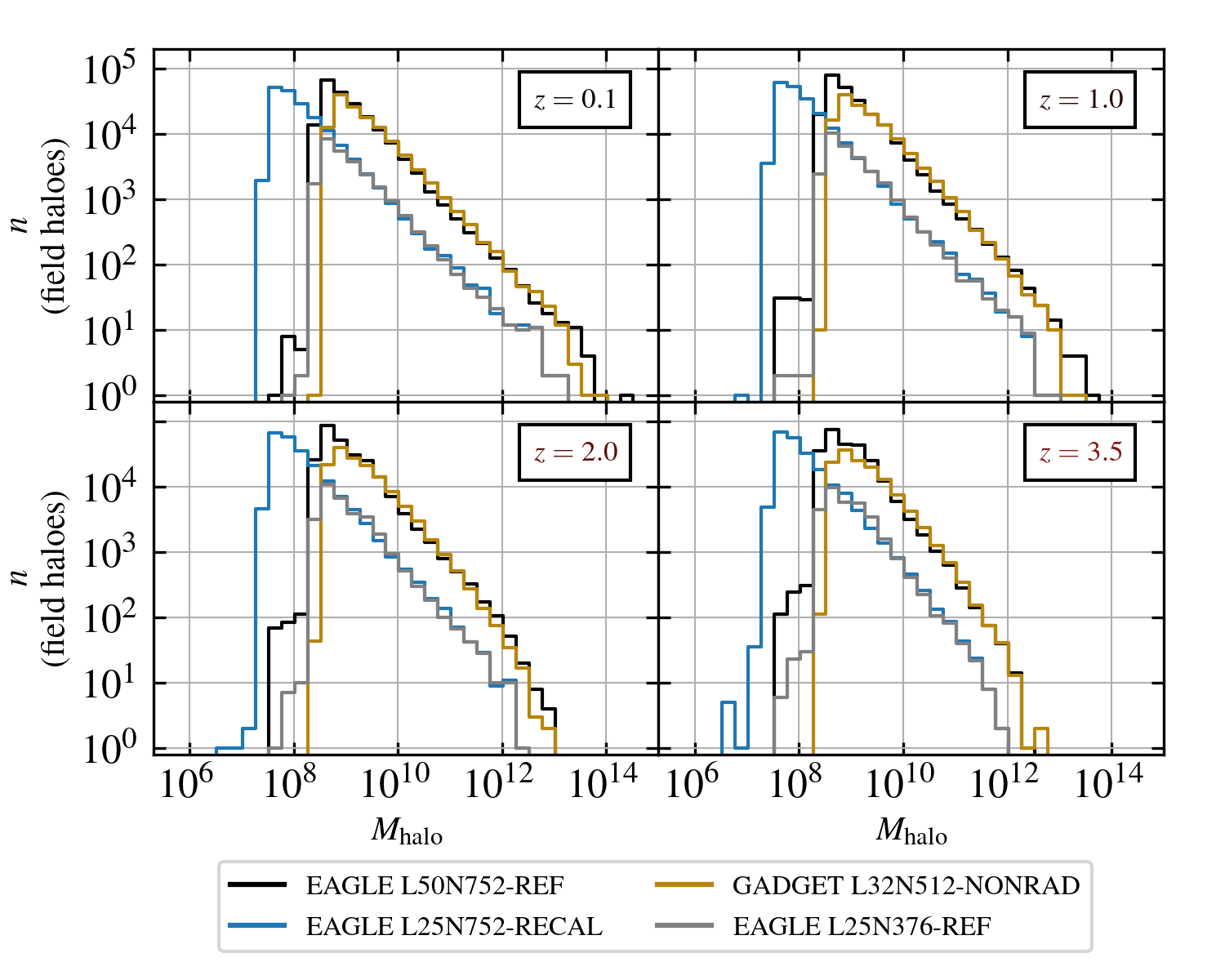}
\caption{The number of field haloes in a selection of runs as a function of halo mass for $4$ redshifts: $z=0.1$ (top left panel), $z=1.0$ (top right panel), $z=2.0$ (bottom left panel), and $z=3.5$ (bottom right panel). In each panel we show data for (i) \eagle\ L50N752-REF (black), (ii) \gadget\ L32N512-NONRAD (khaki), (iii) \eagle\ L25N752-RECAL (blue), and (iv) \eagle\ L25N376-REF (grey). Any alternative physics runs in Table~\ref{tab:sims} with $50$ ($25$) Mpc box size show very similar halo mass distributions as L50-REF (L25-REF). We use $36$ equally log-spaced bins from $M_{\rm halo}=10^{6}M_{\odot}$ to $M_{\rm halo}=10^{15}M_{\odot}$, corresponding to a bin size of $0.25$~dex. }
\label{fig:appendix:massfunctions}
\end{figure*}

With the aim of providing the reader a measure of the sample sizes we use to draw conclusions from, we show in Figure~\ref{fig:appendix:massfunctions} the number of field haloes as a function of halo mass for various runs used throughout the paper. The bin size we use is identical to that used throughout the rest of the paper (unless otherwise stated). 

\section{Resolution convergence}\label{sec:appendix:resolution}
\begin{figure*}
\includegraphics[width=1\textwidth]{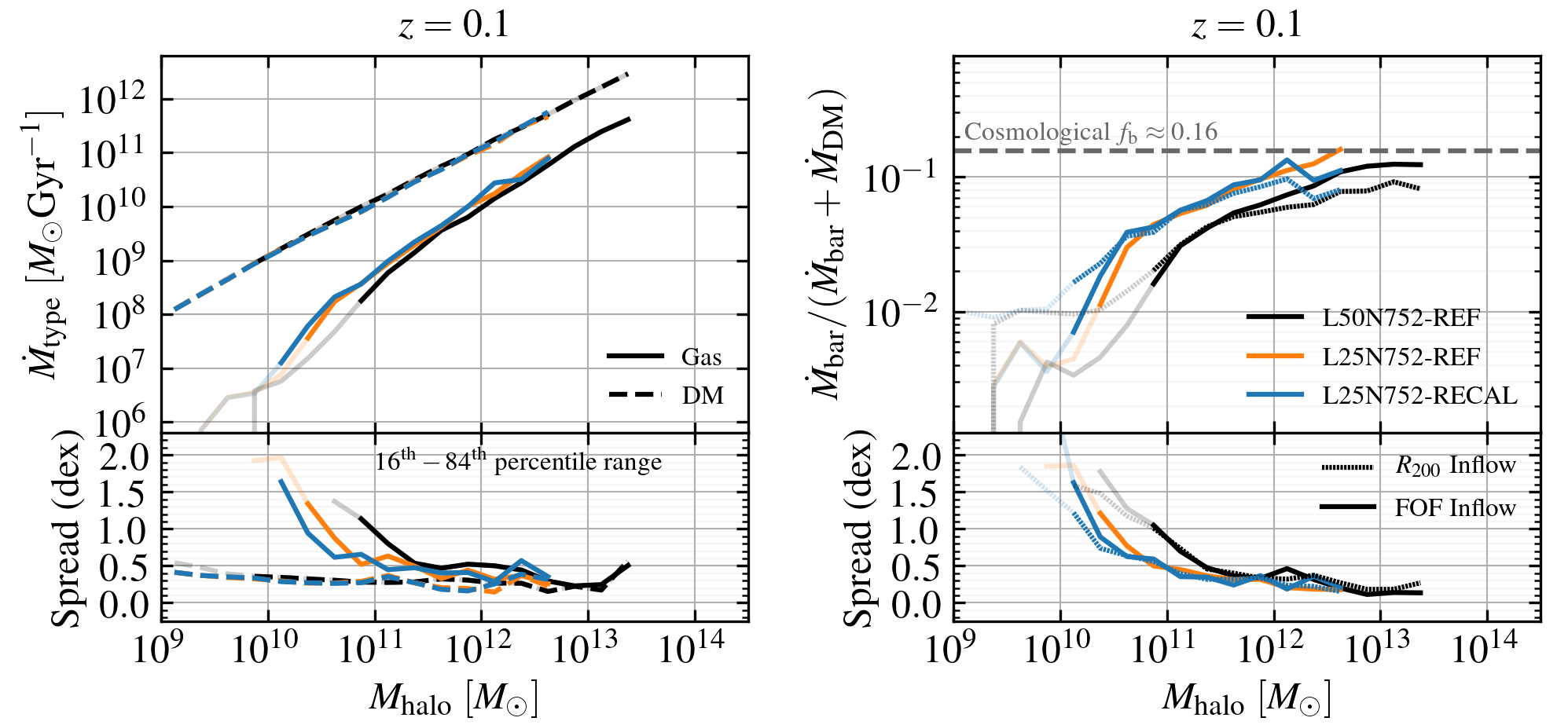}
\caption{Left panels: Inflow rates (top) and spread (bottom) as a function of halo mass for gas (solid lines) and DM (dashed lines), for the standard resolution \eagle\ L50N752-REF box (black) and 2 higher resolution boxes: L25N752-RECAL (blue) and L25N752-REF (orange). Right panels: inflow baryon fractions (top) and spread (bottom) as a function of halo mass, for the same runs. We also compare the baryon fractions from an $R_{200}$ inflow algorithm for the L50N752-REF and L25N752-RECAL runs (densely dotted lines). Line transparency has been increased where the average efficiency has been calculated from a bin in which more than $50$\% of haloes were subject to an accretion flux of less than 50 particles. We see similar results for DM inflow rates irrespective of resolution, and the same trend for gas accretion onto be suppressed at lower halo masses.  We do, however, observe that both higher resolution runs exhibit enhanced gas accretion rates. The spread is marginally reduced for the higher resolution runs at the low mass end. This result is robust using both FOF and $R_{200}$ inflow algorithms.}
\label{fig:appendix:res}
\end{figure*}

\citet{Schaye2015} introduced the concept of ``strong'' and ``weak'' convergence tests. Strong convergence refers to the case where a simulation is re-run at higher resolution, with both better mass and spatial resolution, adopting exactly the same sub-grid physics models and parameters. Weak convergence refers to the case when a simulation is re-run at higher resolution but the sub-grid parameters are recalibrated to recover, as best as possible, the level of agreement with the adopted calibration
diagnostic. In the case of \eagle, the $z=0.1$ galaxy stellar mass
function and stellar size-mass relation of galaxies. With this purpose, two higher-resolution versions of \eagle\ were introduced by \citet{Schaye2015}. Both in a box of
($25$~cMpc)$^{3}$ and with $2\times 752^3$ particles. These simulations have better mass and spatial
resolution than the intermediate-resolution of the L50-REF simulation by factors of
$8$ and $2$, respectively. 
The \eagle\ L25N752-REF adopts identical parameters to the standard resolution reference runs, whereas L25N752-RECAL has $4$ parameters that were recalibrated to reproduce $z=0$ observables above. Hence, a comparison between L25N376-REF and L25N752-REF represents a strong convergence test, while a comparison with L25N752-RECAL is a weak convergence test. Here, we compare with both high resolution runs to examine the convergence of our results.

Figure~\ref{fig:appendix:res} shows accretion rates (left) and inflow baryon fractions (right) as a function of halo mass for the  L25N752-RECAL, L25N752-REF and L25N376-REF (we also include L50-REF in black). Focusing on raw accretion rates (left panels), DM accretion rates agree very well between different resolutions, and show very little spread ($\lesssim0.5$~dex across all halo masses), while gas accretion rates are systematically enhanced in L25N752-REF and L25N752-RECAL compared to L50-REF by $\approx 0.1-0.2$~dex. The natural spread in gas accretion rates increases towards lower halo masses for the 3 runs shown, however the magnitude of the spread at a given halo mass is slightly less in the higher resolution runs. In the right panels we illustrate $f_{\rm b,\ inflow}$ using our primary FOF method (solid lines), and also using the $R_{200}$ spherical overdensity method (densely dotted lines, only for L50-REF and L25N752-RECAL). Similar to raw accretion rates, we see that baryon accretion rates relative to DM are systematically enhanced in L25N752-REF and L25N752-RECAL compared to L50-REF by $\approx0.1-0.2$~dex (the offset fairly steady over halo mass). We see the same enhancement of gas accretion using our $R_{200}$ method, indicating that this result is not a consequence of methodology, and is rather a physically meaningful reflection of the different accretion rates between the simulations. Although enhancement of gas accretion is seen with higher resolution, the decrease in gas accretion relative to DM remains significant nonetheless. We therefore remain confident in our conclusions regarding $f_{\rm b,\ inflow}$ using the standard resolution L50-REF run. Enhanced gas accretion rates in the higher resolution runs offers an explanation for differences between these two simulations previously reported. Figure 11 in \citet{Schaye2015} and Figure B1 in \citet{Collacchioni19} show enhanced specific star formation rates (SSFRs) and accretion rates onto galaxies that leads to star formation, respectively, in the L25N752-RECAL run relative to the $100$~Mpc reference box for galaxies in the stellar mass range $M_{\star}\lesssim 10^{10}M_{\odot}$. These enhancements are likely the consequence of the enhanced baryon inflow rate at the halo level, which increases the abundance of gas available for eventual star formation. Based on the findings in Figure \ref{fig:appendix:res}, we find convincing strong/weak convergence between L50-REF and L25N752-REF/L25N752-RECAL when measuring DM accretion rates, and slight tension when comparing gas accretion rates (the tension being quantitatively consistent when using recalibrated parameters compared to reference parameters).

\section{Temporal convergence}\label{sec:appendix:dt}
\begin{figure}
\includegraphics[width=0.45\textwidth]{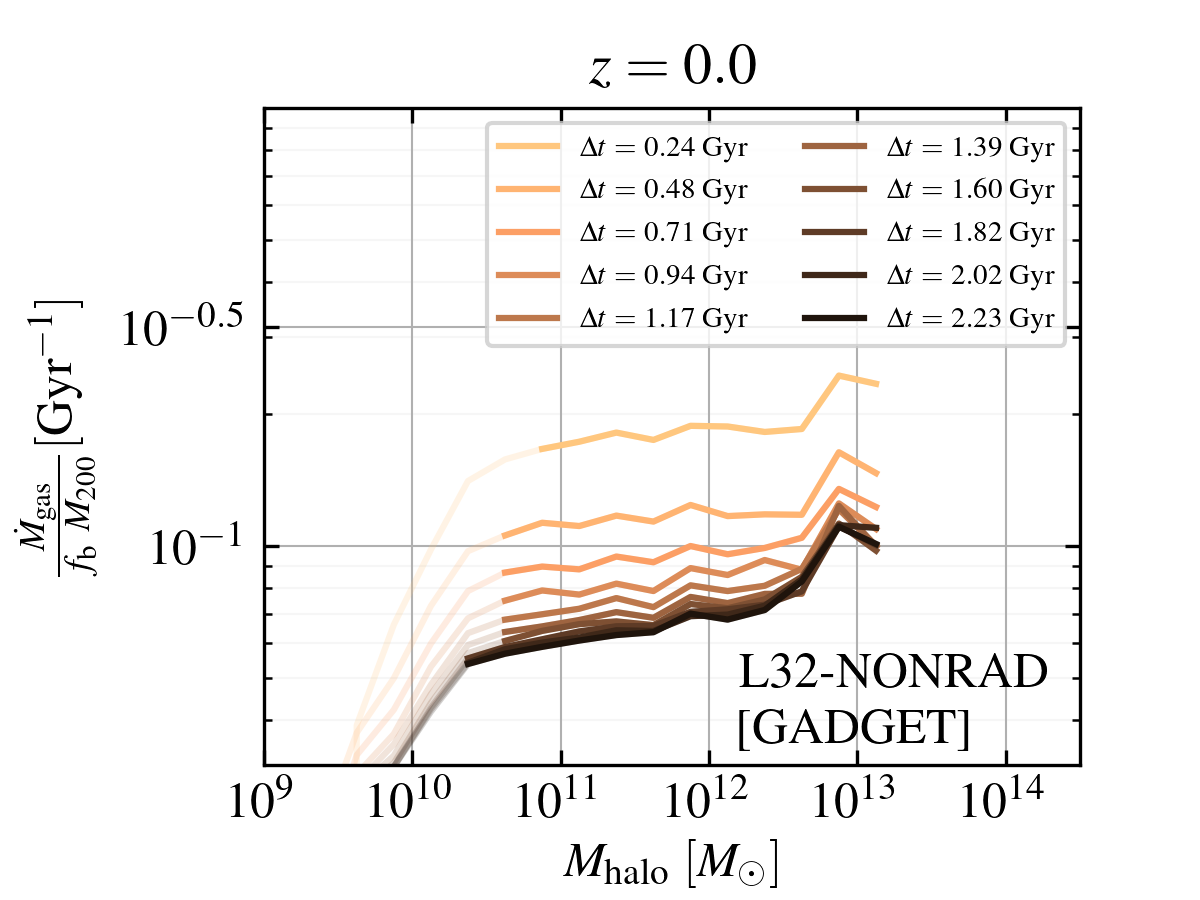}
\caption{Inflow efficiencies (calculated using a \gadget-SPH based non-radiative $32$Mpc box with $512^3$ DM and gas particles) as a function of $M_{200}$ for a collection of $\Delta t$ values (all intervals ending at $z\approx0$). Shorter accretion intervals show increased instantaneous accretion flux, while efficiencies appear to converge towards correct time-integrated inflow flux when we use longer $\Delta t$ values, $\gtrsim1$~Gyr, at $z\approx0$. A similar result is found using full \eagle\ physics in \protect{\citet{Mitchell2020}}. }
\label{fig:appendix:temporalconvergence}
\end{figure}

As explained in \S \ref{sec:methods:chumm}, our accretion rates are calculated over a finite time interval, and our accretion rates are therefore sensitive to the choice of $\Delta t$ . In the case of \eagle, we elect to use adjacent snapshots to constitute the time interval for accretion (corresponding to a $\Delta t$ value closest to $\Delta t\approx1$ Gyr at $z\approx0$, or, more generally,  $\Delta t\approx 0.5-1\times t_{\rm dyn}(z)$). We would consider our results to be converged if our accretion rates represent an accurate time=integrated measure of matter inflow, with minimal contribution from stochastic particle accretion and outflow events over short timescales ($\ll t_{\rm dyn}(z)$). Figure~\ref{fig:appendix:temporalconvergence} shows accretion efficiencies at $z\approx0$ in the L32-NONRAD run as a function of halo mass, for a number of different $\Delta t$ interval lengths to investigate the sensitivity of our algorithm to the time interval. We find with adiabatic physics that accretion rates converge towards longer intervals, with shorter $\Delta t$ intervals leading to higher instantaneous accretion rates. This is likely a consequence of more stochastic inwards crossings of the FOF boundary over shorter timescales, while longer $\Delta t$ intervals yield a more accurate time integrated accretion rate. Given our  temporal convergence test in Figure~\ref{fig:appendix:temporalconvergence} using L32-NONRAD, we can be confident that our calculations in \eagle\ using longer $\Delta t$ intervals have converged towards accurate time integrated accretion rates. This is further evidenced in the convergence tests presented in \citet{Mitchell2020}, who show a similar trend using full \eagle\ physics. It should also be noted that the behaviour of accretion efficiency with halo mass remains identical comparing different $\Delta t$ values, albeit with different normalisation - meaning that even if the interval we used did not yield accurate time-converged accretion rates, our qualitative results would not be significantly influenced. 

\section{FOF inflow compared to SO inflow}\label{sec:appendix:r200}
\begin{figure}
\includegraphics[width=0.45\textwidth]{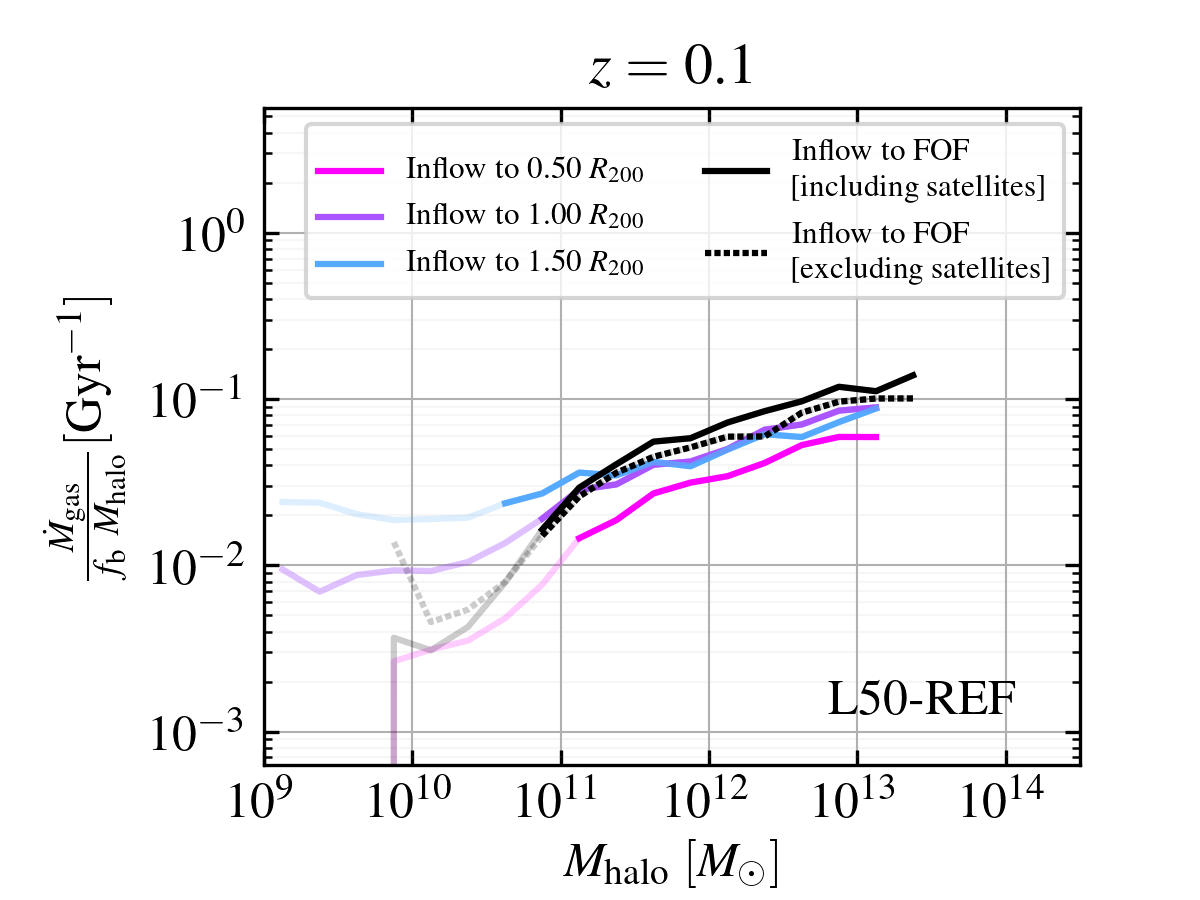}
\caption{Inflow efficiencies at $z\approx0$ for the L50-REF box over a collection of $R_{200}$ fractional spheres (coloured lines), compared to (i) FOF inflow {\it including satellites} (our primary method, black solid line), and (ii) FOF inflow onto  {\it only the central subhalo} (i.e. neglecting satellites, black dotted line). Line transparency has been increased where the average efficiency has been calculated from a bin in which more than $50$\% of haloes were subject to an accretion flux of less than 50 particles. We observe that our FOF algorithm agrees well with the $R_{200}$ algorithm at halo masses below $10^{11.5}M_{\odot}$, but shows slightly enhanced gas accretion above this mass scale, which we attribute to accretion onto satellite subhaloes. When we only include accretion onto the central subhalo, we also see good agreement with the $R_{200}$ calculation to within $0.1$~dex for halo masses above $10^{11.5}M_{\odot}$.}
\label{fig:appendix:r200}
\end{figure}

Figure \ref{fig:appendix:r200} illustrates the behaviour of our primary FOF-based accretion algorithm (solid black line) compared to an $R_{200}$ spherical mass flux based calculation (coloured lines representing inflow over a collection of fractional $R_{200}$ spheres). We also show accretion efficiencies if we neglect to include accretion rates to satellite subhaloes (dotted line) compared to the standard FOF algorithm (which includes accretion onto satellites). We note that the FOF algorithm neglecting satellite accretion agrees very well with accretion flux to the $R_{200}$ sphere (purple line). Unsurprisingly, including accretion onto satellites increases accretion rates, particularly for higher mass haloes ($M_{\rm halo}\gtrsim10^{11.5}M_{\odot}$), which are host to more substructure. This increase compared to $R_{200}$-based inflow is small, of order $\approx0.1-0.2$~dex. This comparison gives us confidence in our results, indicating that the influence of the method is both minimal and predictable. 


\bsp	
\label{lastpage}
\end{document}